\documentclass[12pt]{iopart}

\usepackage{iopams}
\expandafter\let\csname equation*\endcsname\relax
\expandafter\let\csname endequation*\endcsname\relax 
\usepackage{amssymb,amsmath}
\usepackage{graphicx}
\usepackage{geometry}
\usepackage{array}
\usepackage{nomencl}
\usepackage[colorlinks,bookmarksopen,bookmarksnumbered,citecolor=red,urlcolor=red]{hyperref}
\usepackage{setspace}
\usepackage{threeparttable}
\usepackage{stmaryrd}

\newcommand{\rev}[1]{#1}
\newcommand{\complete}[1]{#1}

\newcommand{\el}{h}
\newcommand{\ml}{\ell}
\newcommand{\Ol}{L}

\newcommand{\Eps}{E}
\newcommand{\MEps}{\M{\Eps}}

\newcommand{\eps}{\varepsilon}
\newcommand{\Meps}{\M{\eps}}

\newcommand{\Mdiff}{\M{\partial}}

\newcommand{\edge}[1]{\multicolumn{1}{||c||}{#1}}
\newcommand{\ledge}[1]{\multicolumn{1}{||c|}{#1}}
\newcommand{\redge}[1]{\multicolumn{1}{|c||}{#1}}
\newcommand{\M}[1]{{\boldsymbol #1}}
\newcommand{\T}{T} 
\newcommand{\Aref}[1]{\ref{#1}}
\newcommand\de[1]{\,{\mathrm d}#1}
\newcommand{\trn}{^{\sf T}}
\newcommand{\puc}{\mathcal{Y}}
\newcommand{\stress}{\sigma}
\newcommand{\Msig}{\M{\sigma}}
\newcommand{\sigfl}[2]{\stress_{#1}^{*(#2)}}
\newcommand{\rsigfl}[2]{\rec{\stress}_{#1}^{*(#2)}}

\newcommand{\set}[1]{{\mathbb #1}}
\newcommand{\Mgfune}{\M{\Gamma}^{0}} 
\newcommand{\N}{\set{N}}
\newcommand{\R}{\set{R}}
\newcommand{\x}{\M{x}}
\newcommand{\y}{\M{y}}
\newcommand{\p}{\M{p}}
\newcommand{\Ps}{\set{P}}
\newcommand{\half}{\mbox{$\frac{1}{2}$}} 
\newcommand{\nc}{n^\textrm{\ec}}
\newcommand{\nt}{n^\textrm{t}}
\newcommand{\nNW}{n^\textrm{NW}}
\newcommand{\ncs}{n^\textrm{cs}}
\newcommand{\rd}{\rho}
\newcommand{\dmn}{\mathcal{O}}
\newcommand{\dmnS}{\mathcal{O}_\mathrm{S}}
\newcommand{\dmnT}{\mathcal{O}_\mathrm{\T}}
\newcommand{\SII}{S_2}
\newcommand{\vfrac}{\phi}
\newcommand{\Zdmn}{\set{K}^\dmn}
\newcommand{\ZdmnS}{\set{K}^{\dmnS}}
\newcommand{\ZdmnT}{\set{K}^{\dmnT}}

\newcommand{\dmnN}{n^\dmn}
\newcommand{\dmnNS}{n^{\dmnS}}
\newcommand{\dmnNT}{n^{\dmnT}}

\renewcommand{\k}{\M{k}}
\newcommand{\cF}{\chi}
\newcommand{\m}{\M{m}}
\newcommand{\Wa}[3]{W#1/#2--#3}
\newcommand{\rec}[1]{\widetilde{#1}} 

\newcommand{\rS}{\rec{S}_2}

\newcommand{\f}{f}
\newcommand{\fS}{\f^\mathrm{S}}
\newcommand{\fSigloc}[1]{\fSig_{#1}}
\newcommand{\fSig}{\f^{\Sigma}}

\newcommand{\fT}{\f^\mathrm{\T}}
\newcommand{\wf}{w}

\newcommand{\ec}{c}
\newcommand{\numGe}[1]{n^{\Ge{#1}}}

\newcommand{\np}{n^\mathrm{\id}}
\newcommand{\iW}{t}
\newcommand{\new}[1]{\widehat{#1}}

\newcommand{\npedge}{n^\mathrm{\id}}
\newcommand{\conf}[2]{#1\{#2\}}
\newcommand{\U}{U}
\newcommand{\tmp}{\theta}
\newcommand{\tmpmax}{\tmp^{\max}}
\newcommand{\tmpmin}{\tmp^{\min}}
\newcommand{\tmpmult}{\tmp^\mathrm{mlt}}
\newcommand{\nmax}{n^{\max}}
\newcommand{\id}{d}
\newcommand{\ef}{q}

\newcommand{\Ap}{A^\mathrm{d}}
\newcommand{\E}[1]{\Omega_{#1}}
\newcommand{\G}{\Gamma}
\newcommand{\Ge}[1]{\G_{#1}}

\newcommand{\uflc}[2]{u_#1^{*(#2)}}
\newcommand{\lc}[1]{^{(#1)}}
\newcommand{\nno}{n^\mathrm{n}}
\newcommand{\MN}{\M{N}}
\newcommand{\Ma}{\M{a}}
\newcommand{\Mu}{\M{u}}

\newcommand{\flc}{^*}
\newcommand{\glb}{^0}

\newcommand{\MU}{\M{\Psi}}
\newcommand{\Mb}{\M{b}}
\newcommand{\MS}{\M{\Sigma}}
\newcommand{\MT}{\M{\T}}
\newcommand{\rTfl}[2]{\rec{\T}_{#1}\flc{}\lc{#2}}
\newcommand{\rMTfl}{\rec{\M{\T}}\flc}
\newcommand{\jmp}[1]{[ #1 ]}

\newcommand{\MTfl}{\MT\flc}
\newcommand{\Mnu}{\M{\nu}}
\newcommand{\Tfl}[2]{\T_{#1}\flc{}\lc{#2}}

\newcommand{\Ges}[2]{\M{\G}_{#1,#2}}

\newcommand{\Ed}{E^\mathrm{\id}}
\newcommand{\Em}{E^\mathrm{m}}
\newcommand{\nud}{\nu^\mathrm{\id}}
\newcommand{\num}{\nu^\mathrm{m}}
\newcommand{\avg}[1]{{\langle #1 \rangle}}
\newcommand{\ML}{\M{L}}
\newcommand{\npuc}{n^\puc}

\begin{document}

\title{Microstructural enrichment functions based on stochastic Wang tilings}

\author{Jan Nov\'{a}k}
\address{Institute of Structural Mechanics, Faculty of Civil Engineering, Brno
University of Technology, Veve\v{r}\'{\i} 95, 602 00 Brno, Czech Republic}
\address{Department of Mechanics, Faculty of Civil Engineering, Czech Technical University in Prague, Th\'{a}kurova 7, \mbox{166 29 Praha 6}, Czech Republic}
\ead{novakj@cml.fsv.cvut.cz}
\author{Anna Ku\v{c}erov\'{a}}
\address{Department of Mechanics, Faculty of Civil Engineering, Czech Technical University in Prague, Th\'{a}kurova 7, \mbox{166 29 Praha 6}, Czech Republic}
\ead{anicka@cml.fsv.cvut.cz}
\author{Jan Zeman}
\address{Department of Mechanics, Faculty of Civil Engineering, Czech Technical University in Prague, Th\'{a}kurova 7, \mbox{166 29 Praha 6}, Czech Republic}
\address{Centre of Excellence IT4Innovations, V\v{S}B-TU Ostrava, 17. listopadu
15/2172, 708 33 Ostrava-Poruba, Czech Republic}
\ead{zemanj@cml.fsv.cvut.cz}

\begin{abstract}
This paper presents an approach to constructing microstructural enrichment
functions to local fields in non-periodic heterogeneous materials with
applications in Partition of Unity and Hybrid Finite Element schemes. It is
based on a concept of aperiodic tilings by the Wang tiles, designed to produce
microstructures morphologically similar to original media and enrichment
functions that satisfy the underlying governing equations. An appealing feature
of this approach is that the enrichment functions are defined only on a small
set of square tiles and extended to larger domains by an inexpensive stochastic
tiling algorithm in a non-periodic manner. Feasibility of the proposed
methodology is demonstrated on constructions of stress enrichment functions
for two-dimensional mono-disperse particulate media.
\end{abstract}


\section{Introduction}\label{sec:introduction}
\rev{%
A detailed analysis of microstructured materials 
with the full resolution of heterogeneities by classical finite element
methods has been found computationally
prohibitive~\cite{oden2006revolutionizing}. To overcome this, one option
consists of modelling a coarse-scale problem with the help of homogenisation
techniques based on effective material
properties~\cite{Feyel2000309,pichler2008micron,Geers2010}. However, this may
lead to a considerable loss of information on the fine scale behaviour, thereby
resulting in an inaccurate assessment of microstructural effects on the
global response and/or its evolution.

An alternative, computationally appealing, strategy proceeds from generalised
finite element formulations that enhance the approximation properties of
standard finite element spaces by subscale-informed enrichment functions. Their
design involves two related but contradictory aspects: (i)~realistic
representation of the underlying heterogeneity patterns and (ii)~construction of
complex enrichment functions in a computationally efficient manner. Here, we
briefly review these issues for two finite element frameworks. The first one is
based on the \emph{partition of unity method}, introduced by Melenk and
Babu\v{s}ka~\cite{Melenk1996289} and generalised in numerous aspects later
on~\cite{Belytschko:2009:REG,Fries:2010:EGF}. The second one utilises the
\emph{hybrid Trefftz stress formulations} developed by Teixeira de
Freitas~\cite{Freitas1998}, see also~\cite{herrera2000trefftz} for an overview.
For simplicity, we restrict our attention to the small-strain linear elasticity
in two dimensions. The following nomenclature is used in the sequel. Scalar
quantities are denoted by plain letters, e.g. $a$ or $A$, vectors and matrices
are in bold as, e.g. $\M{a}$ or $\M{A}$. In addition, we adopt the Mandel
vector-matrix representation of symmetric second- and fourth-order tensors, e.g.
$a_{ij}$ or $A_{ijkl}$, so that~\cite[Section~2.3]{Milton:2002:TC}
\begin{align*}
\M{a} 
=
\begin{bmatrix}
a_{11} \\ a_{22} \\ \sqrt{2} a_{12} 
\end{bmatrix}
, &&
\M{A}
=
\begin{bmatrix}
A_{1111} & A_{1122} & \sqrt{2} A_{1112} \\
A_{2211} & A_{2222} & \sqrt{2} A_{2212} \\
\sqrt{2} A_{1211} & \sqrt{2} A_{1222} & 2A_{1212} 
\end{bmatrix}.
\end{align*}
\nomenclature{$\M{a}$, $\M{A}$}{A matrix or vector}%

\begin{figure}[ht]
  \centering
  \begin{tabular}{ccc}
    \includegraphics[height=35.5mm]{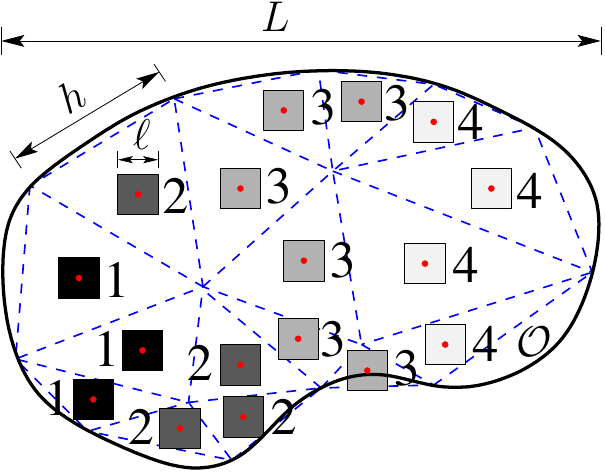} &
    \includegraphics[height=35.5mm]{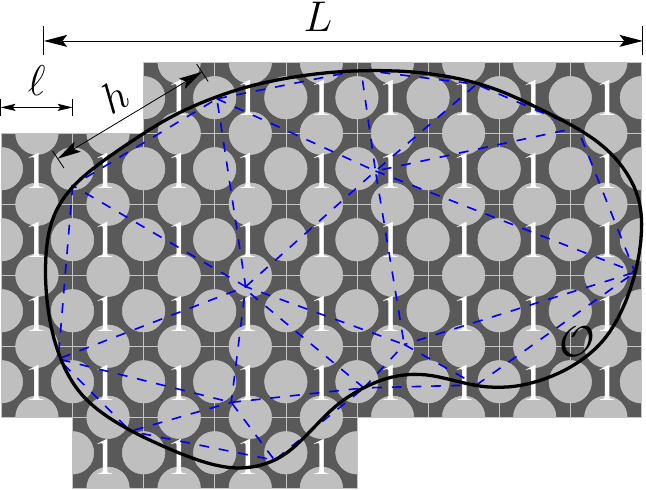} &
    \includegraphics[height=35.5mm]{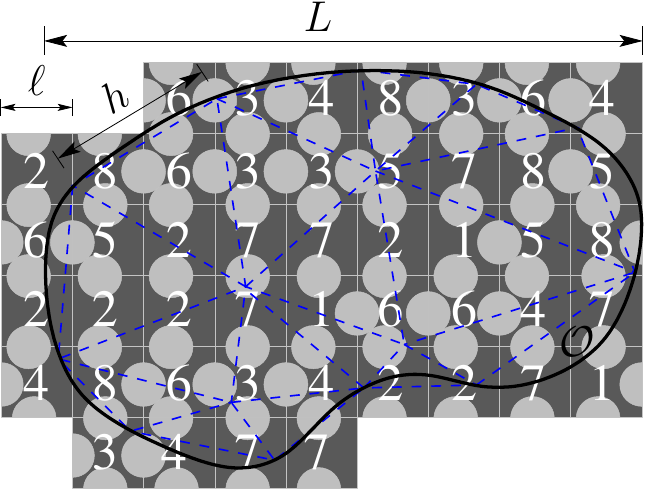} \\
    (a) & (b) & (c)
  \end{tabular}
  \caption{%
  Examples of heterogeneity representations for a macroscopic body $\dmn$ and
  (a)~separated scales ($\ml \ll \el < \Ol$): three unit cells associated with
  integration points (in red), (b) periodic geometry with non-separated scales
  ($\ml < \el < \Ol$): a single periodic unit cell, (c) aperiodic geometry with
  non-separated scales ($\ml < \el < \Ol$): eight distinct Wang tiles;
  $\ml$, $\el$, and $\Ol$ denote characteristic macroscopic, element
  (meso-scopic), and heterogeneity lengths, respectively.}
  \label{fig:domains}
\end{figure}
\nomenclature{$\dmn$}{Domain}%
\nomenclature{$\ml$}{Microstructure length}%
\nomenclature{$\Ol$}{Macroscopic length}%
\nomenclature{$\el$}{Element length}%

\subsection{Partition of unity methods}

Consider a microstructured two-dimensional domain $\dmn \subset \R^2$
approximated by finite elements, cf.~\Fref{fig:domains}.
\nomenclature{$\set A$}{A set}%
\nomenclature{$\set R$}{Set of real numbers}%
The partition of unity methods build on the displacement field approximation in
the form
\begin{equation}\label{eq:partion_of_unity}
\Mu(\x)
\approx
\sum_{n = 1}^{\nno}
N_n(\x)
\left[  
\Ma_n 
+
\MU\flc( \x ) 
\Mb_n
\right]
\mbox{ for }
\x \in \dmn,
\end{equation}
\nomenclature{$a\flc$}{Fluctuating part of $a$}%
\nomenclature{$\Ma$}{Standard degrees of freedom}%
\nomenclature{$\Mb$}{Extended degrees of freedom}%
\nomenclature{$\MU$}{Matrix of enrichment functions}%
\nomenclature{$N$}{Nodal basis functions}%
where $\nno$ is the number of nodes in the finite element mesh, $N_n : \dmn
\rightarrow \R$ denotes the standard finite element basis functions and $\Ma_n
\in \R^2$ the regular degrees of freedom associated with the $n$-th node,
whereas $\MU\flc$ and $\Mb_n$ designate the strategy-specific matrices of
enrichment functions and extended degrees of freedom, respectively.
The ansatz~\eqref{eq:partion_of_unity} is then employed in the standard Galerkin
procedure to arrive at a system of linear(ized) equations involving both
regular and extended degrees of freedom. This approach was explored by Fish and
Yuan~\cite{Fish:2005:MEB,Fish:2007:MEB}, who derived the enrichment functions
from solutions to a periodic unit cell problem, formulated for cells associated
with integration points, see \Fref{fig:domains}(a). In particular,
\begin{equation}\label{eq:displ_enrichments}
\MU\flc(\x) 
=
\begin{bmatrix}
\uflc{1}{1} & \uflc{1}{2} & \frac{1}{\sqrt 2}\uflc{1}{3} \\
\uflc{2}{1} & \uflc{2}{2} & \frac{1}{\sqrt 2}\uflc{2}{3} 
\end{bmatrix}
(\x),
\end{equation} 
\nomenclature{$\uflc{i}{j}$}{the $i$-th component of the fluctuating
displacement field due to the $j$-th impulse}%
where $\uflc{i}{j} : \dmn \rightarrow \R$ denotes the $i$-th component of the
fluctuating displacement field, determined for a unit cell subject to the
average strain with the $j$-th component set to one, while the remaining two
vanish (see \ref{app:mech_fields} for further details). Such form of
enrichment functions is motivated by the displacement decomposition
\begin{equation}
\Mu(\x) = \Mu\glb(\x) + \Mu\flc(\x)
\mbox{ for }
\x \in \dmn,
\end{equation}
with $\Mu\glb:\dmn \rightarrow \R^2$ and $\Mu\flc:\dmn \rightarrow \R^2$
referring to global and fluctuating displacement fields; parameter $\Mb_n \in
\R^3$ in \Eref{eq:partion_of_unity} has thus the physical meaning of a
generalised average strain known from classical homogenisation
theories~\cite{Geers2010}. Since such fields are constructed under the
assumption of separated lenghtscales, \Fref{fig:domains}(a) with $\ml/\Ol
\rightarrow 0$, an attention is paid neither to the geometrical compatibility
among neighbouring cells, nor to the compatibility of the corresponding
enrichment fields. Consistent mathematical results for \emph{periodic media}
with a finite ratio $\ml/\Ol$, \Fref{fig:domains}(b), were obtained by
Matache~\etal\cite{Matache:2000:GPFEM}. The enrichment functions are
constructed on the basis of the spectral version of the unit cell
problem~\cite{Morgan:1991:AAC} resolved by the $p$-version of the finite element
method, see~\cite{Babuska:2011:OLA} for additional contributions to this
field.

The partition of unity methods have also been applied to simulations of material
systems with \emph{explicitly represented non-periodic} heterogeneities, such as
thin fibres~\cite{Radtke:2010:PUM,Radtke:2011:PUM}. Here, the enrichment
function is chosen to be piecewise constant in fibre and matrix domains, and the
extended degrees of freedom correspond to a relative slip at the fibre-matrix
interface. Such simple format comes at the expense of the fact that two extra
degrees of freedom are introduced per fibre, which renders realistic simulations
costly.

\subsection{Trefftz method}

The hybrid Trettfz approach has recently been employed
by Nov\'{a}k~\etal\cite{Novak:2012:MEF} to simulate composites reinforced with
non-periodic ellipsoidal heterogeneities with non-separated lengthscales. The
method builds on the additive stress decomposition
\begin{equation}
\Msig(\x)
=
\Msig\glb(\x)
+
\Msig\flc(\x)
\mbox{ for }
\x \in \dmn,
\end{equation}
\nomenclature{$\stress$}{Stress}%
with $\Msig\glb : \dmn \rightarrow \R^3$ corresponding to the macroscopic stress
field and $\Msig\flc : \dmn \rightarrow \R^3$ being stress fluctuations,
approximated at the level of an element $\E{e}$ as
\nomenclature{$\E{e}$}{$e$-th finite element}%
\begin{equation}\label{eq:trefftz_approximation}
\Msig(\x)
\approx
\MS_e(\x)
\Ma_e
+
\MS\flc(\x)
\Mb_e
\mbox{ for }
\x \in \E{e}.
\end{equation}
\nomenclature{$\MS$}{Stress basis functions}%
Here, in analogy to \Eref{eq:partion_of_unity}, $\MS_e : \E{e} \rightarrow \R^{3
\times m}$ stands for the standard basis functions of the Trefftz method associated
with $m$ regular degrees of freedom $\Ma_e \in R^m$ and $\Mb_e \in \R^3$ denotes
the extended degrees of freedom with the physical meaning of average element
strains. The individual enrichment functions
\begin{equation}\label{eq:stress_enrichments}
\MS\flc( \x )
=
\begin{bmatrix}
\sigfl{11}{1} & \sigfl{11}{2} & \sqrt{2} \sigfl{11}{3} \\
\sigfl{22}{1} & \sigfl{22}{2} & \sqrt{2} \sigfl{22}{3} \\
\sqrt{2} \sigfl{12}{1} & \sqrt{2} \sigfl{12}{2} & 2 \sigfl{12}{3} 
\end{bmatrix}( \x )
\mbox{ for }
\x \in \dmn,
\end{equation}
correspond to the fluctuating stress fields due to unitary strain
impulses, see again \ref{app:mech_fields} for further details. Note that the
regular and enrichment basis functions need to be selected such that the stress
remains self-equilibrated. The stress approximation is complemented with an
independent approximation of displacements at the element boundary
$\G_e$~\cite{Kaczmarczyk20091298}
\nomenclature{$\G$}{Boundary or edge}%
\begin{equation}
\Mu(\x)
\approx
\MN^\G_e( \x )
\Ma^\G_e
\mbox{ for }
\x \in \G_e,
\end{equation}
\nomenclature{$\MN$}{Matrix of basis functions}%
involving only regular edge shape functions $\MN^\G_e$ and regular boundary
degrees of freedom $\Ma^\G_e$. The remainder of the formulation follows from the
weak form of the equilibrium and compatibility equations, which can be converted
to the element boundaries by virtue of the divergence theorem,
cf.~\cite{Kaczmarczyk20091298,Freitas1998}. The appealing feature of the
particular formulation~\cite{Novak:2012:MEF} is that the size of the resulting
system of equations is the same as for the homogeneous problem, due to the
elimination of the extended degrees of freedom. This is achieved by a careful
construction of the enrichment functions through Eshelby solutions for
individual particles~\cite{eshelby1957determination,Eshelby:1959:EFO}, combined
together to obtain compatible mechanical fields~\cite{Novak:2012:MEF}.

\subsection{Tiling-based approach}

This short overview illustrates the major difficulty in simulating non-periodic
systems with realistic geometries, namely that simple enrichment functions lead
to the loss of information and/or to a significant increase in the number of
degrees of freedom, whereas manageable system sizes necessitate complex
constructions of enrichment functions. The aim of this work is thus to develop
an algorithm that allows for extending the local (possibly periodic)
data from computationally tractable samples to entire macroscopic domains
in a non-periodic way, \Fref{fig:domains}(c). The algorithm keeps the
synthesised enrichment functions, $\MU\flc$ in \Eref{eq:partion_of_unity} or
$\MS\flc$ in \Eref{eq:trefftz_approximation}, continuous across congruent
boundaries and consistent in terms of statistical properties of original and
reconstructed material morphologies. It is based on a small number of the
so-called Wang
tiles~\cite{wang1961tiling,glassner2004andrew,Culik:1996:ASW:245761.245814} and
a stochastic tiling procedure introduced by Cohen et al.~\cite{cohen2003wang}.

In 1961, Hao Wang introduced a tiling concept involving square tiles with
different codes on their edges, referred to as Wang tiles~\cite{wang1961tiling}.
The tiles are connected together so that the adjacent edges have the same code
and permit a computationally efficient graphic reproduction of morphological
patterns~\cite{cohen2003wang,Culik:1996:ASW:245761.245814,demaine2007jigsaw,glassner2004andrew}.
Their desirable aesthetic properties are attributed to the aperiodicity of
tilings, whereas the low computational effort results from the use of a small
number of tiles to compress the entire morphological
information~\cite{lagae2008comparison}.

Here, we exploit and extend these principles to provide a basis for an efficient
generation of microstructure-based enrichment functions applicable in partition
of unity or hybrid Trefftz finite element algorithms. In
order to meet additional criteria arising from such constructions, the Simulated
annealing-based optimisation~\cite{Kirkpatrick:1983:S,Cerny:1985:JOTA} is used
to arrive at optimal tile sets. The performance of the method is illustrated on
the construction of tile-based stress enrichment functions in a mono-disperse
two-phase composite medium with linear elastic phases. Although the proposed
approach is illustrated solely in the two-dimensional setting, it is fully
extensible to three dimensions by exploring the results available for the Wang
cubes~\cite{Culik:1995:ASW,Lu:2007:VIW}. We also note in passing that the
techniques developed in this paper can be used equally well as microstructure
reconstruction or generation algorithms, generalising the previous developments
available e.g.
in~\cite{Povirk:1995:IMI,Yeong:1998:RRM,Kumar:2006:UMR,Zeman:2007:FRM,Lee:2009:3DR,Niezgoda2010:OMB,Schoder:2011:ARM}.
}

The paper structure is as follows. The concept of stochastic Wang tiling is
described in \Sref{s:tiling}. A discussion on the optimisation procedure based
on prescribed statistical descriptors and compatibility of synthesised
mechanical fields on contiguous tile edges is given in
\Sref{s:microstructure_optimization}. \Sref{s:results} comprises numerical
examples demonstrating the performance of the proposed approach. Final remarks
on the current developments and future plans are assembled
in~\Sref{s:conclusions}. Finally, in \Aref{app:mech_fields}, we present a brief
overview of the stress analysis algorithm utilised to determine the local stress
fluctuations.

\section{Aperiodic tilings by sets of Wang tiles}\label{s:tiling}

\begin{figure}[ht]
  \centering
  \begin{tabular}{ccc}
    (a)~
    \includegraphics[height=55mm]{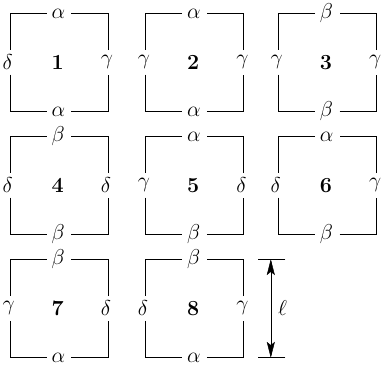} 
    &&
    (b)~
    \includegraphics[height=55mm]{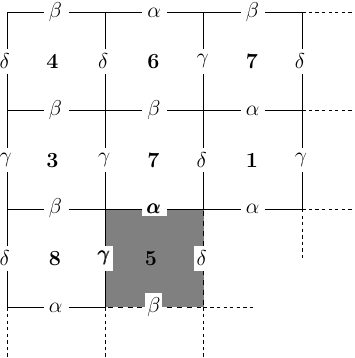}
    \end{tabular}
  \caption{(a)~Tile set W8/2-2~\cite{cohen2003wang} consisting of $8$ tiles with
  $2$ vertical $\{\alpha, \gamma\}$ and $2$ horizontal $\{ \beta, \delta \}$ edge
  codes with equal frequencies of occurrence
  $\ef_\alpha = \ef_\beta = \ef_\gamma = \ef_\delta =
  \frac{1}{4}$ , and $\nNW = 2$, (b) an example of aperiodic valid
  tiling with highlighted connectivity across south-eastern and north-western edges.}
  \label{fig:wang_tile}
\end{figure}

Consider again the domain $\dmn$ \rev{from \Fref{fig:domains}(c)} covered by a
regular square grid. Each grid cell contains specific microstructural patterns
that are compatible on contiguous boundaries. If there are no missing cells
inside the synthesised domain, the discretization is called a valid
\emph{tiling}\footnote{
Henceforth, the term ``tiling'' stands for ``valid tiling'' exclusively, thereby
excluding invalid tilings from the consideration.}  and a single cell is
referred to as the \emph{Wang tile}~\cite{wang1961tiling},
\Fref{fig:wang_tile}. The tiles have different codes on their edges,
enumerated here by lowercase Greek letters, and are not allowed to rotate during
the tiling procedure. The number of distinct tiles is fixed, though arranged
in such a fashion that no sub-sequence of tiles periodically repeats.
The set of all distinct tiles is referred to as the \emph{tile set}, \Fref{fig:wang_tile}(a).
Sets that enable uncountably many, always aperiodic, tilings are called
\emph{aperiodic} sets~\cite{Culik:1996:ASW:245761.245814}. The assumption of
strictly aperiodic sets can be relaxed, though still being capable to tile the
plane aperiodically, e.g., when utilising the Cohen-Shade-Hiller-Deussen~(CSHD)
tiling algorithm~\cite{cohen2003wang} briefly introduced in the following
section. Note that such tilings provide substantial generalisations to periodic
paving algorithms, which use identical tiles--periodic unit cells, \rev{recall
\Fref{fig:domains}(b)}.

\subsection{Tile set setup}

Favourable properties of a tile set to control repetitive effects
proceed from the tile and edge code diversity. The number of edge
codes $\nc_i$ in the $i$-th spatial direction of the Cartesian coordinates can
be chosen arbitrarily, while the number of tiles $\nt$ must satisfy
\nomenclature{$\nc$}{Number of edge codes}%
\nomenclature{$\nt$}{Number of tiles}%
\nomenclature{$\nNW$}{Number of NW tiles}%
\nomenclature{$\ncs$}{Number of tiles in the complete set}%
\begin{equation}\label{eq:n^t_us}
\nt = \nNW \sqrt{\ncs},
\end{equation}
where $\ncs = (\nc_1 \nc_2)^2$ is the number of tiles in the complete set and
$\nNW = 2, \ldots, \sqrt{\ncs}$ stands for the number of tiles
associated with each admissible pair of north-western~(NW) edge codes,
\Fref{fig:wang_tile}(a), see~\cite{Novak:2012:CRM} for further details.

When designing a tile set, one chooses a particular number of edge codes $\nc_1$
and $\nc_2$. The complete set of $\ncs$ tiles is created by mutually permuting
the codes. In order to tile the plane, the south-eastern edge codes must match
those assigned to NW edges, \Fref{fig:wang_tile}(b). Thus, the created tiles are
collected according to NW combinations. Finally, a desired number of tiles
is chosen using \Eref{eq:n^t_us}, \rev{in such a way that $\nNW$ unique
tiles is selected from each NW group}. The emerging, user-defined, set of tiles
is referred to as \Wa{$\nt$}{$\nc_1$}{$\nc_2$}. Moreover, we denote the relative
frequency of occurrence of the $\ec$-th code in the tile set by $\ef_\ec$, see
\Fref{fig:wang_tile}(a).
\nomenclature{\Wa{$x$}{$y$}{$z$}}{Wang tile set composed of $x$ tiles and $y$
and $z$ edge codes}%
\nomenclature{$\ef$}{Frequency of occurrence}%

\subsection{CSHD stochastic tiling algorithm}\label{sec:stoch_tiling_algorithm}

Since there are $\nNW$ tiles associated to each NW group, index of the new tile
to be placed is selected randomly from the set $\{ 1, \ldots, \nNW \}$ with the
uniform probability. Beforehand, one must select an appropriate NW group
compatible with the eastern code of a previously placed tile and the southern
code of the tile just above the one to be placed (edges $\alpha$ and $\gamma$ of
shaded areas in~\Fref{fig:wang_tile}(b)). Aperiodicity of the resulting tiling
is guaranteed by assuming that the random generator never returns a periodic
sequence of numbers and that each NW group contains at least two distinct
tiles~\cite{cohen2003wang}.

\section{Designing optimal tile set morphology}\label{s:microstructure_optimization}
%
To simplify the exposition, we limit our attention to two-phase composite
media formed by a matrix phase and equi-sized disks of radius $\rd$ and
a parametric microstructure representation built on the Wang tile set
\Wa{8}{2}{2}\footnote{%
\nomenclature{$\rd$}{Disk radius}%
The set~\Wa{8}{2}{2} has been chosen since it is the simplest one that allows
for aperiodic patterns in the stochastic sense~\cite{cohen2003wang}. Note that
all the steps of the tile set design can be directly generalised to more complex
tile sets, cf.~\cite{Novak:2012:CRM}.}, introduced in
\Sref{sec:microstructure_parameterization}.
The location of the disks within the tiles has to be optimised to achieve
(i)~good approximation of the original microstructure in terms of a given
morphological descriptor, \Sref{sec:stat_prop}, and (ii)~microstructures that
guarantee the compatibility of \rev{enrichment functions} on contiguous tile
edges,
\Sref{sec:objective_stress}. Such criteria originate from different
perspectives. The first goal aims at capturing the dominant spatial features of
original media, while the latter criterion ensures that the tiling-generated
fields comply with the governing differential equations. \rev{The details of the
algorithm used to solve the resulting optimisation problem are provided in
\Sref{sec:optimization_procedure}.}

\subsection{Microstructure parametrisation}\label{sec:microstructure_parameterization}
The adopted \rev{bitmap-based} microstructure representation \rev{involves a}
Wang tile set consisting of $\nt$ tiles of the edge length $\ml \in \N$~(in
pixels), in which we distribute $\np$ disks of radius $\rd$. The $\id$-th disk
is represented by a triplet $\{ \iW_\id, x_{1,\id}, x_{2,\id} \}$, where
$\iW_{\id} \in \{1, \ldots, \nt \}$ denotes the tile index and $x_{\id,j} \in
\{1, \ldots, \ml \}$ specifies the position of the $\id$-th disk within the tile
at the $j$-th direction. The associated parameter vector $\p$ is obtained as a
collection of these data:
\begin{equation}\label{eq:parameter_def}
\p 
=
\left[
\iW_\id, 
x_{1,\id},
x_{2,\id}
\right]_{\id=1}^{\np}.
\end{equation}
Since the position of each disk is specified by three parameters, the
parameter space~$\Ps$ is $(3 \times \np)$-dimensional, i.e. $\Ps \subset
\N^{3\times\np}$.
\nomenclature{$\np$}{Number of disks}%
\nomenclature{$\iW$}{Index used to number Wang tiles}%
\nomenclature{$\p$}{Parameter vector in optimisation}%
\nomenclature{$\id$}{Parameter used to index disks}%
\nomenclature{$\Ps$}{Parameter set}%

In an admissible configuration, the disks do not penetrate each other or overlap
corners of tiles being associated with. The first constraint reflects the given feature
of the original microstructure, \Fref{fig:original_microstructure}(a), whereas
the latter one arises as an artifact intrinsic to the edge-based tiling
algorithm, e.g.,~\cite{cohen2003wang}. In addition, to maintain the
morphological compatibility, any disk intersecting the edge of a given code
needs also be associated to tiles containing the same edge. To emphasise
this, we encode a particular microstructural configuration as
\conf{$\np$}{$\npedge_\ec$}$_{\ec=1}^{\nc}$, where $\npedge_\ec$ denotes the
number of disks intersecting the edge of code~$\ec$,
see~\Fref{fig:admissible_configuration} on
page~\pageref{fig:admissible_configuration}.
\nomenclature{$\conf{a}{\npedge}$}{Configuration of $a$ disks, in which $\npedge$
intersect edges}

\subsection{Statistical properties of the microstructure}\label{sec:stat_prop}
The most common class of statistical descriptors embodies a set of $n$-point
probability functions, applicable to generic heterogeneous
media~\cite{Torquato:2002}. In this paper, the focus is on the two-point
probability function, which captures primary phenomena as the phase volume
fraction, characteristic microstructural length(s), and long-range
orientation orders, if any.

\begin{figure}[ht]
  \centering
  \begin{tabular}{cc}
    (a)~\includegraphics[height=45mm]{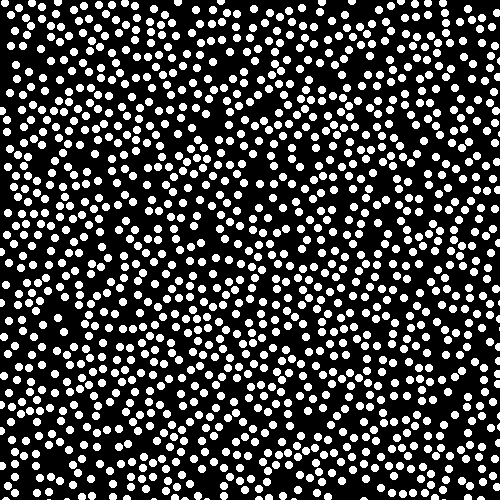}
    &
    (b)~\includegraphics[height=45mm]{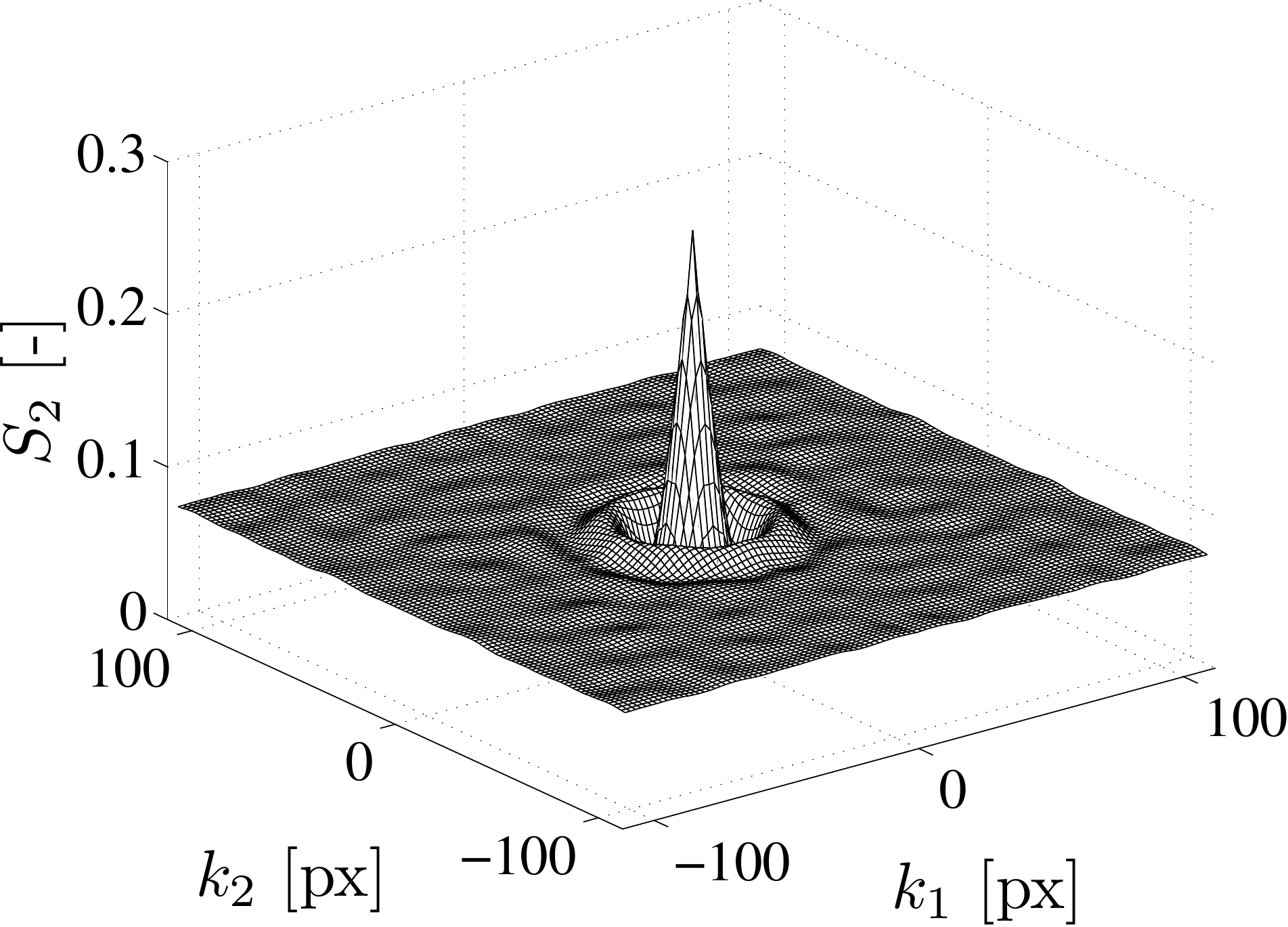}
    \end{tabular}
  \caption{(a) An example of two-phase medium formed by equilibrium distribution
  of $1,300$ equi-sized disks of volume fraction $26.8 \%$ and (b) the
  two-point probability function $\SII$; the sample is discretized with $1,000
  \times 1,000$ pixels and each disk has the radius of $8$ pixels.}
\label{fig:original_microstructure} 
\end{figure}

We now assume that the domain $\dmn$ is occupied by a two-phase heterogeneous
material discretized by a regular lattice \rev{of} $\dmnN_1 \times \dmnN_2$
pixels, indexed by $\k \in \Zdmn$ with
\begin{equation}\label{eq:lattice_def}
\Zdmn
= 
\left\{
\m \in \set{Z}^2 
: 
-\frac{\dmnN_i}{2} < m_i \leq \frac{\dmnN_i}{2}
, 
i = 1, 2
\right\}.
\end{equation}
\nomenclature{$\dmnN_i$}{Number of pixels in $\dmn$ and $i$-th direction}%
\nomenclature{$\Zdmn$}{Set of pixel indices}%
\nomenclature{$\k$}{Pixel index}%
\nomenclature{$\m$}{Auxiliary pixel index}%
The distribution of individual phases (disks and matrix) within $\dmn$ is
quantified by the characteristic function $\cF(\k)$, which equals $1$ when $\k$
is occupied by the disk phase and $0$ otherwise, cf.
\Fref{fig:original_microstructure}(a). Assuming a periodic\footnote{%
Note that periodicity is considered here for the sake of computational
efficiency. The tiling-generated data is always aperiodic.}
ergodic medium, the two-point probability function $\SII : \Zdmn \rightarrow
[0,1]$ is then defined as~\cite{Torquato:2002}
\nomenclature{$\cF$}{Characteristic function}
\nomenclature{$\SII$}{Two-point probability function}
\begin{equation}\label{eq:Sdef}
\SII(\k)
=
\frac{1}{\dmnN_1 \dmnN_2}
\sum_{\m \in \Zdmn}
\cF(\m)
\cF\left(
\left\lfloor \k + \m \right\rfloor_{\Zdmn} \right),
\end{equation}
where $\lfloor \bullet \rfloor_{\Zdmn}$ denotes the $\Zdmn$-periodic extension.
Noticing that~\eqref{eq:Sdef} has the structure of circular correlation, the
two-point probability function can be efficiently evaluated using Fast Fourier
Transform techniques, see e.g.~\cite{Gajdosik:2006:QAFC}.

According to its definition, $\SII(\k)$ quantifies the probability that two
arbitrary points separated by $\k$ will both be located at the disk phase when
randomly selected from $\Zdmn$. Denoting by $\vfrac$ the disk volume fraction,
$0 \leq \vfrac \leq 1$, the two-point probability function satisfies
$\SII(\M{0}) = \vfrac$. Moreover, $\SII(\k) \simeq \vfrac^2$ for $\| \k \| \gg
\rd$ indicates that the medium does not exhibit repeating long-range order
orientation effects, cf.~\Fref{fig:original_microstructure}(b).
\nomenclature{$\vfrac$}{Volume fractions}

\begin{figure}[ht]
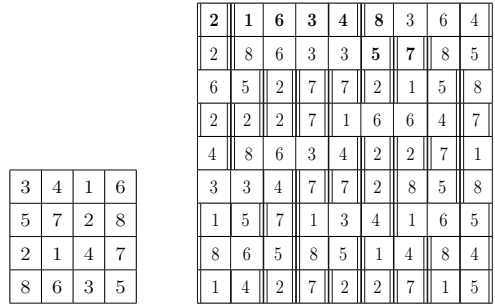

  \centering
  \begin{tabular}{cc}
    \resizebox{1.78cm}{1.78cm}{
    \begin{tabular}[b]{|c|c|c|c|}
      \hline
     3  &   4  &   1  &   6  \\
     \hline
     5  &   7  &   2  &   8  \\
     \hline
     2  &   1  &   4  &   7  \\
     \hline
     8  &   6  &   3  &   5  \\
     \hline
     \end{tabular}}
    &\resizebox{4cm}{4cm}{
    \begin{tabular}[b]{|c|c|c|c|c|c|c|c|c|}
      \hline
     \edge{\bf 2} &   \ledge{\bf 1}  &  \bf 6  &  \bf 3  &  
     \redge{\bf 4} & \ledge{\bf 8} & 3 &   6  &   \redge{4} \\
     \hline
     \edge{2}  &   \ledge{8}  &   6  &   3  &   3  &   \redge{\bf 5}  &  
     \edge{\bf 7} &   \ledge{8}  &   \redge{5} \\
     \hline
     6  &   \redge{5}  &   \edge{2}  &   \edge{7}  &   \edge{7}  &   \edge{2}  &   \ledge{1}  &   \redge{5}  &   \ledge{8} \\
     \hline
     \edge{2}  &   \edge{2}  &   \edge{2}  &   \edge{7}  &   \ledge{1}  &   6  &   6  &   \redge{4}  &   \edge{7} \\
     \hline
     \redge{4}  &   \ledge{8}  &   6  &   3  &   \redge{4}  &   \edge{2}  &   \edge{2}  &   \edge{7}  &   \ledge{1} \\
     \hline
     3  &   3  &   \redge{4}  &   \edge{7}  &   \edge{7}  &   \edge{2}  &   \ledge{8}  &   \redge{5}  &   \ledge{8} \\
     \hline
     \ledge{1}  &   \redge{5}  &   \edge{7}  &   \ledge{1}  &   3  &   \redge{4}  &   \ledge{1}  &   6  &   \redge{5} \\
     \hline
     \ledge{8}  &   6  &   \redge{5}  &   \ledge{8}  &   \redge{5}  &   \ledge{1}  &   \redge{4}  &   \ledge{8}  &   \redge{4} \\
     \hline
     \ledge{1}  &   \redge{4}  &   \edge{2}  &   \edge{7}  &   \edge{2}  &   \edge{2}  &   \edge{7}  &   \ledge{1}  &   \redge{5} \\
     \hline
     \end{tabular}}\\
    (a) Tiling $\dmnS$ &(b) Tiling $\dmnT$
  \end{tabular}
  \caption{Valid tilings used in optimisation with respect to (a)~two-point
  probability function and (b) stress field; highlighted vertical
  edges in (b)~correspond to edge set $\Ge{\delta}$ containing $50$
  equivalent edges of code $\delta$ and \rev{length $\ml$. Tiles denoted by
  bold numbers~(first two rows) are used to generate aperiodic enrichment
  functions.}}
\label{fig:optimproc}
\end{figure}

The following procedure is adopted to determine the two-point probability
function for the tile-based microstructure. First, the set \Wa{8}{2}{2} is used
to assemble a $4 \times 4$ tiling $\dmnS \subset \R^2$, \rev{periodic on
external boundaries}, in which each tile appears with the same frequency in
order to suppress artificial fluctuations in volume fractions,
\Fref{fig:optimproc}(a). The domain $\dmnS$ is discretized by an $\dmnNS_1
\times \dmnNS_2$ regular grid with the same pixel size as in the original
microstructure, so that $\dmnNS_i < \dmnN_i$. Given a parameter vector $\p$
quantifying positions of individual disks, the tile-based morphology is
quantified by the two-point probability function $\rS : \Ps \times \ZdmnS
\rightarrow [0,1]$, and its proximity to the target microstructure is evaluated
as
\begin{equation}\label{eq:obj_func_tppF}
\fS( \p )
=
\frac{1}{\dmnNS_1 \dmnNS_2}
\sum_{\k \in \ZdmnS}
\left( 
\SII(\k)
-
\rS(\p, \k)
\right)^2,
\end{equation}
where $\ZdmnS$ is defined analogously as for the target medium
$\dmn$.
\nomenclature{$\fS$}{Effect function based on two-point probability}
\nomenclature{$\dmnNS$}{Number of lattice points for tiling}
\nomenclature{$\rS$}{Two-point probability function of tiling}
\nomenclature{$\ZdmnS$}{Indexes for two-point probability function}

\renewcommand{\arraystretch}{1.3}
\renewcommand{\arraystretch}{1}

\rev{%
\subsection{Stress-based enrichment fields and their
compatibility}\label{sec:objective_stress}

The additional, yet more complex, goal is to find the tile set morphology that
ensures the admissibility of enrichment functions synthesised by the tiling
algorithm. Analogously to the original Wang idea, this is achieved by requiring
that edges of identical codes carry identical, this time non-scalar,
information. In particular, motivated by encouraging results obtained recently
in~\cite{Novak:2012:MEF}, we concentrate on the stress enrichment
functions $\MS\flc$, recall \Eref{eq:stress_enrichments}. It is natural
convert them to equivalent traction fluctuations, obtained as
\nomenclature{$\nu_i$}{Normal vector component}%
\begin{equation}\label{eq:traction_enrich}
\MTfl
=
\Mnu
\MS{}\flc
,\;
\end{equation}
\complete{where $\MTfl$ collects the components associated
with individual load-cases and $\Mnu$ stores the components of the normal
vector:}
\begin{align*}
\MTfl
=
\begin{bmatrix}
\Tfl{1}{1} & \Tfl{1}{2} & \Tfl{1}{3} \\ 
\Tfl{2}{1} & \Tfl{2}{2} & \Tfl{2}{3}
\end{bmatrix}
, &&
\Mnu
=
\begin{bmatrix}
\nu_1 & 0 & \frac{1}{\sqrt 2} \nu_2 \\
0 & \nu_2 & \frac{1}{\sqrt 2} \nu_1
\end{bmatrix}.
\end{align*}
\nomenclature{$\Mnu$}{Matrix of normal vector components}%
\nomenclature{$\Tfl{i}{j}$}{$i$-th components of tractions due to $j$-th%
loadcase}%

Analogously to the morphology design, the definition of the traction-based
objective function is based on an auxiliary $9 \times 9$ tiling $\dmnT$,
\Fref{fig:optimproc}(b), discretized into $\dmnNT_1 \times \dmnNT_2$ bitmap with
pixels indexed by $\k \in \ZdmnT$. The tiling is periodic at external
boundaries, and contains all admissible combinations of tile pairs from the set
\Wa{8}{2}{2} sharing all edge codes\footnote{There are $16$ distinct pair
combinations of basic tiles sharing the code $\delta:~\{ 2-1, 2-2, 2-7, 2-8,
4-1, 4-2, 4-7, 4-8, 5-1, 5-2, 5-7, 5-8, 7-1, 7-2, 7-7, 7-8 \}$ see
\Fref{fig:wang_tile}(a). All these combinations are present in the tiling
$\dmnT$ in \Fref{fig:optimproc}(b), each of them multiple times.}, since we assume that the
edge traction values are dominated by the response of adjacent tiles. Hence, for
each edge code $\ec \in \{ 1, 2, \ldots, \nc \}$ with $\nc = \nc_1 + \nc_2$, we
introduce a set $\Ge{\ec}$ formed by $\numGe{\ec}$ edges  of identical code
$\ec$ and length $\ml$ with normal vector $\Mnu^{\Ge{\ec}}$,
\Fref{fig:optimproc}(b)\footnote{For
the particular tile set considered here, we set $\Mnu^{\Ge{\alpha}} =
\Mnu^{\Ge{\gamma}} = [0,1]$ and $\Mnu^{\Ge{\beta}} = \Mnu^{\Ge{\delta}} =
[1,0]$.}. By $\Ges{c}{j} : \{1, \ldots, \ml\} \rightarrow \ZdmnT$, $j \in \{ 1, 2, \ldots, \numGe{\ec} \}$, we denote a
function providing coordinates of individual pixels at the $j$-th edge of
code~$c$.

\nomenclature{$\dmnT$}{Tiling for traction objective function}%
\nomenclature{$\ZdmnT$}{Indices for traction-based objective function}%
\nomenclature{$\ec$}{Edge code}%
\nomenclature{$\Ge{\ec}$}{All edges of code $\ec$}%
\nomenclature{$\Ges{c}{j}$}{Coordinates of $j$-edge of code $\ec$}%

Now we are in a position to quantify differences of tractions carried by an edge
code $\ec$, due to differing neighbours, via an objective function $\fT_\ec$.
For a given parameter vector $\p \in \Ps$ and material properties of individual
phases, we calculate the stress enrichment function $\MS\flc( \p, \k )$ by the
algorithm outlined in \Aref{app:mech_fields}, and evaluate the objective
function as
\begin{equation}
\fT_\ec(\p)
=
\frac{1}{\ml}
\sum_{s=1}^\ml
\left\| 
\max\left\{
\MTfl( \p, \Ges{c}{j}(s))
\right\}_{j=1}^{\numGe{\ec}}
-
\min\left\{
\MTfl( \p, \Ges{c}{j}(s))
\right\}_{j=1}^{\numGe{\ec}}
\right\|_1,
\end{equation}
\nomenclature{$\fT$}{Traction-based objective function}%
where the traction enrichments are determined from~\Eref{eq:traction_enrich}
with $\Mnu = \Mnu^{\Ge{\ec}}$, $\max$ and $\min$ operations are understood
component-wise and $\| \M{A} \|_{1} = \sum_{i,j} | A_{ij} |$. Collecting the
contributions from all codes, we obtain
\begin{equation}\label{eq:obj_func_tractions}
\fT(\p)
=
\sum_{\ec=1}^{\nc}
\fT_\ec(\p).
\end{equation}

Once the tile set is designed with respect to the objective
function~\eqref{eq:obj_func_tractions}, the tiling-based stress enrichment
functions $\rec{\MS}\flc : \ZdmnT \rightarrow \R^{3\times 3}$ can be assembled
by the CSHD algorithm using the stress fluctuations $\MS\flc$ carried by an
arbitrary selection of tiles $1$--$8$ from the tiling $\dmnT$. In the numerical
experiments reported in \Sref{s:results}, we use the set of eight tiles from the
top rows of $\dmnT$ highlighted by bold numbers in \Fref{fig:optimproc}(b), but
equivalent results were obtained for different selections. Due to this
procedure, the reconstructed edge tractions corresponding to the synthesised
enrichments $\rec{\MS}\flc$ may experience jumps at tile edges. For the $j$-th
edge of the set $\Ge{\ec}$, these are defined as
\begin{equation}\label{eq:trct_jmp}
\jmp{\rMTfl_{\ec,j}}(s)
=
\Mnu^{\Ge{\ec}}
\left(
\rec{\MS}\flc_{+}(\Ges{\ec}{j})
-
\rec{\MS}\flc_{-}(\Ges{\ec}{j})
\right)
\mbox{ for }
s \in \{1, 2, \ldots, \ml \},
\end{equation}
where $\rec{\MS}\flc_{+}$ and $\rec{\MS}\flc_{-}$ denote the values of the
stress enrichment functions taken from the nearest edge neighbours from right
and left, respectively, relative to the orientation of the edge set $\Ge{\ec}$
by the normal vector $\Mnu^{\Ge{\ec}}$.
\nomenclature{$\rec{\MS}\flc$}{Reconstructed stress enrichment functions}%
}

\subsection{Optimisation procedure}\label{sec:optimization_procedure}

In fact, the goals represented by objective functions~\eqref{eq:obj_func_tppF}
and~\eqref{eq:obj_func_tractions} are conflicting. Minimising only with respect
to the two-point probability function results in \rev{traction enrichments
discontinuous at internal edges}, whereas the latter criterion drives the system
to a periodic distribution of disks. To achieve a compromise solution, we
introduce a composite objective function in the form 
\begin{equation}\label{eq:err_tot}
\f(\p)
=
\wf 
\fS(\p)
+
\fT(\p),
\end{equation}
where $\wf$ denotes a weighting factor \rev{balancing geometrical features with
mechanical compatibility}.
\nomenclature{$\f$}{General objective function}%
\nomenclature{$\wf$}{Weighting factor}%
The minimisation of the objective function~\eqref{eq:err_tot} is performed by
the well-established Simulated Annealing
method~\cite{Kirkpatrick:1983:S,Cerny:1985:JOTA}, extended by a re-annealing
phase to escape from local extremes, e.g.~\cite{Leps2005PhD}.

\begin{figure}[h]
\centerline{%
\includegraphics[height=40mm]{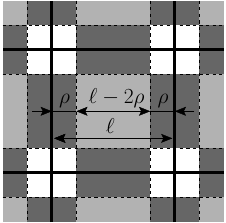}
}
\caption{Tile decomposition into interiors, edges and corner
regions.}
\label{fig:tile_decomposition}
\end{figure}

\begin{figure}[ht]
  \centering
  \includegraphics[width=\textwidth]{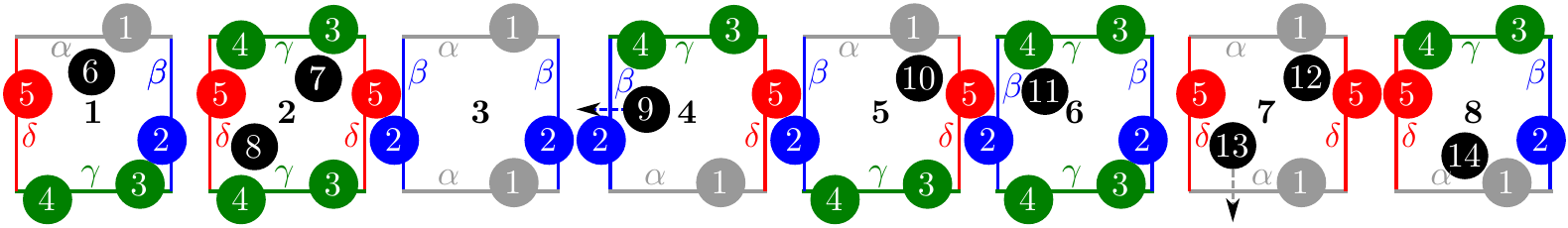}
  \caption{An example of an admissible \conf{14}{1-1-2-1}
  configuration~(with $\npedge_\alpha = \npedge_\gamma =
  \npedge_\delta = 1$ and $\npedge_\beta=2$ code-related disks) and its
  modification by disk displacements; disk 9 leaves its parent tile 4 and
  randomly enters tiles 1, 3, 6 or 8; disk 13 leaves its parent tile 7 and
  randomly enters tiles 1, 3, 5 or 7.}
  \label{fig:admissible_configuration} 
\end{figure}

Given the number of disks $\np$ and the target volume fraction $\vfrac$, we
initiate the algorithm by determining the number of edge disks $\npedge_\ec$
related to the $\ec$-th code and the tile edge length $\ml$. Although this
problem is difficult due to multiplicity of the edge-related disks, recall
\Fref{fig:admissible_configuration}, we resolved it by a heuristic
procedure outlined next. To this purpose, an arbitrary tile is
decomposed into three regions assigned to interiors~(light grey area in
\Fref{fig:tile_decomposition}), edges~(dark grey area in
\Fref{fig:tile_decomposition}), and to corners~(white area in
\Fref{fig:tile_decomposition} that cannot be occupied by disks due to the
corner constraint). For a disk configuration
$\conf{\np}{\np_\ec}_{\ec=1}^{\nc}$ related to a tile set
\Wa{$\nt$}{$\nc_1$}{$\nc_2$}, there is $(\np -
\sum_{\ec=1}^{\nc} \np_c)$ interior disks and, due to the edge constraints,
a single disk associated with code $\ec$ appears $2\nt q_\ec$ times,
cf.~\Fref{fig:admissible_configuration}. Thus, 
the disk volume fraction in the tile set or in a tiling is given by
\begin{equation}\label{eq:global_volume_fraction}
\rec{\vfrac}
\approx
\frac{\Ap}{\nt \ml^2}
\left( 
\np 
+ 
\sum_{\ec=1}^{\nc}
\left(
2\nt \ef_\ec - 1  
\right) \npedge_c
\right),
\end{equation}
\nomenclature{$\Ap$}{Area of a disk}%
\nomenclature{$\rec{\vfrac}$}{Reconstructed volume fraction}%
with $\Ap$ denoting the area of a single disk~(in square pixels), and should be
as close to the target value $\vfrac$ as possible. In addition, we impose the
condition
\begin{equation}\label{eq:local_volume_fraction}
\frac{\np 
- 
\sum_{\ec=1}^{\nc} \npedge_c }{(\ml - 2 \rd)^2}
\approx
\frac{\nt \sum_{\ec = 1}^{\nc} \ef_\ec \npedge_\ec}{2\rho( \ml - \rd )},
\end{equation}
matching the local volume fractions of disks in interior and edge regions. 
Thus, given the numbers of disks attached to codes $\{
\npedge_\ec \}_{\ec=1}^{\nc}$, Eqs.~\eqref{eq:global_volume_fraction}
and~\eqref{eq:local_volume_fraction} implicitly define tile edge lengths
$\tilde{\ml}$ and $\hat{\ml}$, which should be equal to each other for
the correct tile set setup. In our case, we sequentially check all values $\{
\npedge_\ec\}_{c=1}^{\nc}$ such that $\npedge_\ec \geq 0$, 
$\sum_{\ec=1}^{\nc} \npedge_\ec \leq \np$ and select the
configuration with the minimum difference $|\tilde{\ml} - \hat{\ml} |$.\footnote{%
Note that the values of $\ml$ and $\np$ are kept constant during the
optimisation process, whereas the values of $\npedge_\ec$ are allowed to change,
since disks can move freely between tile interiors and edges.}
\nomenclature{$\npedge_\ec$}{Number of disks associated with edges of code
$\ec$}%
\nomenclature{\conf{$a$}{$b$}}{Configuration of $a$ disks with $b$
disks intersecting edges}

On the basis of these data, we randomly generate positions of individual disks
and assign them to randomly selected tile interiors and edges, until
an admissible configuration $\p$ is obtained. A single loop of the optimisation
algorithm involves a sequential selection of a disk $\id \in \{1, \ldots, \np
\}$, and its movement given by
\begin{align}\label{eq:particle_move}
\new{x}_{j,\id}
= 
x_{j,\id} 
+ 
\ml 
\left( 
  \U 
  - 
  \half 
\right),
&&
j = 1, 2,
\end{align}
repeated until a new admissible configuration $\new{\p}$ is encountered. The
symbol $U$ denotes a random variable with a uniform distribution in the interval
$[0,1]$. If a disk, during its displacement, leaves its parent tile by crossing
the edge of code $\ec$, it is randomly assigned to a tile sharing the same
code, \Fref{fig:admissible_configuration}.
\nomenclature{$\U$}{Random variable with a uniform distribution on [0,1]
interval}

The acceptance of the new solution $\new{\p}$ is driven by the Metropolis
criterion~\cite{Kirkpatrick:1983:S}
\begin{equation}\label{eq:Metropolis_criterion}
\exp \left(
\frac{\f(\p) - \f(\new{\p})}{\tmp}
\right)
\geq \U, 
\end{equation}
where $\tmp$ denotes the algorithmic temperature, initially set to $\tmpmax$ and
gradually reduced by a constant multiplicator $\tmpmult < 1$ once the loop over
all $\np$ disks is completed. The entire algorithm terminates after $\nmax$
objective function evaluations. Moreover, we keep it restarting when the current
temperature $\tmp$ is less than the threshold value $\tmpmin$. Such a
re-annealing step was found beneficial, as the resulting problem is multi-modal
and discontinuous due to the presence of edge-constrained disks.
\nomenclature{$\tmp$}{Algorithmic temperature}
\nomenclature{$\tmpmax$}{Maximum temperature}
\nomenclature{$\tmpmin$}{Minimum temperature}
\nomenclature{$\tmpmult$}{Temperature multiplication}
\nomenclature{$\nmax$}{Maximum number of function evaluations}

\section{Results}\label{s:results}
The potential of the tile-based representation is demonstrated for the two-phase
composite medium appearing in \Fref{fig:original_microstructure}, \rev{with
default parameters shown in~\Tref{tab:default_parameters}}. Distinct sets
\Wa{8}{2}{2}, differing in (i)~the tile edge length $\ml$, (ii)~the number of
total and edge disks \conf{$\np$}{$\npedge_\ec$}$_{\ec=1}^{\nc}$, (iii)~the
weighting factor $\wf$, and in~\rev{(iv) phase properties contrast $\Ed/\Em$}
have been examined. In particular, our aim is to demonstrate that the proposed
tile morphology design procedure works well and that the
tile sets based on the specific tilings $\dmnS$ and $\dmnT$ can be used to
represent generic particulate media.

\begin{table}[ht]
\centering
\begin{threeparttable}
\caption{Default setting of parameters.}
\label{tab:default_parameters}
\begin{tabular}{lr}
\hline
\multicolumn{2}{c}{\bf{Microstructure}}\\
\hline
Volume fraction, $\vfrac$ & $26.8\%$ \\
Disk radius, $\rd$ & $8$~pixels \\
Young modulus of disk\tnote{a}, $\Ed$ & 10 \\
Young modulus of matrix\tnote{a}, $\Em$ & 1 \\
Poisson ratio of matrix and disks, $\num = \nud$ & $0.125$ \\
\hline
\multicolumn{2}{c}{\bf{Optimisation algorithm}}\\
\hline
Weighting factor\tnote{b}, $w$ & $10^5$ \\
Maximum temperature, $\tmpmax$ & $10^{-3}$ \\
Minimum temperature, $\tmpmin$ & $10^{-6}$ \\
Multiplicative factor, $\tmpmult$ & $(\tmpmax / \tmpmin)^{1/200}$ \\
Number of function evaluations, $\nmax$ & $10^4\np$ \\
\hline
\end{tabular}
\begin{tablenotes}
\item[a] In what follows, all stress-related values are expressed in consistent
units.

\item[b] Determined as $\wf \approx \avg{\fT} / \avg{\fS}$, with e.g.
$\avg{\fS}$ denoting the average value of $\fS$ determined for $20$ randomly
generated disk configurations.

\end{tablenotes}
\end{threeparttable}
\end{table}
\nomenclature{$\Ed$, $\Em$}{Young modulus of disk and matrix}%
\nomenclature{$\nud$, $\num$}{Poisson ratio of disk and matrix}%
\nomenclature{$\avg{\bullet}$}{Average value of $\bullet$}%

\begin{figure}[ht]
  \centering
  \begin{tabular}{cc}
    (a)~\conf{10}{1-1-0-0}, $\rec{\vfrac} = 23.6\%$ & 
    \conf{10}{1-0-1-1}, $\rec{\vfrac} = 28.0\%$  \\
    \includegraphics[width=4cm]{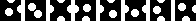} &
    \includegraphics[width=4cm]{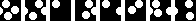} 
    \\
    (b)~\conf{17}{1-1-1-1}, $\rec{\vfrac} = 27.9\%$,  
    & \conf{17}{1-1-1-1}, $\rec{\vfrac} = 27.9\%$ \\
    \includegraphics[width=4.96cm]{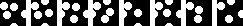} &  
    \includegraphics[width=4.96cm]{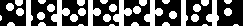} \\
    (c)~\conf{27}{2-1-1-1}, $\rec{\vfrac} = 26.7\%$ & 
    \conf{27}{2-1-1-1}, $\rec{\vfrac} = 26.7\%$ \\
    \includegraphics[width=6.08cm]{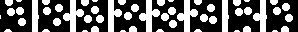} &
    \includegraphics[width=6.08cm]{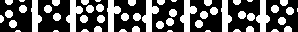} \\ 
    (d)~\conf{32}{2-1-2-1}, $\rec{\vfrac} = 26.6\%$ & 
    \conf{32}{1-2-2-1}, $\rec{\vfrac} = 26.6\%$ \\
    \includegraphics[width=7.04cm]{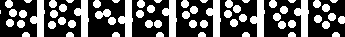} & 
    \includegraphics[width=7.04cm]{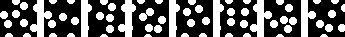} \\
    $\wf = 10^4$ & $\wf = 10^5$ \\
    \end{tabular}
  \caption{Optimised sets \Wa{8}{2}{2} obtained for weighting
  factors $\wf$ equal to $10^4$ and $10^5$ and for configurations with
  (a)~$\np = 10$, $\ml = 42$~px, 
  (b)~$\np = 17$, $\ml = 52$~px, 
  (c)~$\np = 27$, $\ml = 64$~px and
  (d)~$\np = 38$, $\ml = 74$~px; 
  \conf{$\np$}{$\npedge_\ec$}$_{\ec=1}^{\nc}$ refers to configuration of $\np$
  disks in total with $\npedge_\ec$ disks intersecting edge $\ec$, $\ml$
  is the tile edge length and $\rec{\vfrac}$ is the reconstructed volume
  fraction.}
\label{fig:optimal_tile_sets} 
\end{figure}

\begin{figure}[ht]
  \centering
  \begin{tabular}{ccc}
    (a) &
    \includegraphics[height=5cm]{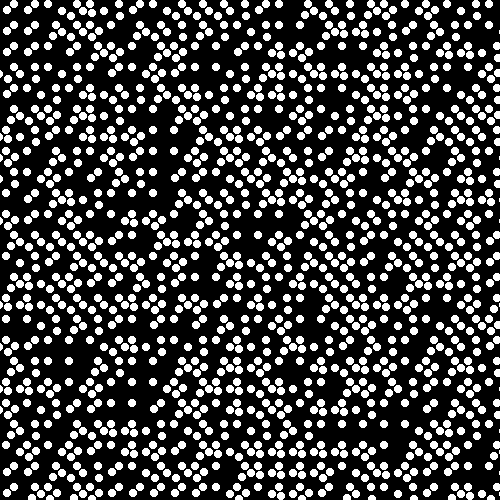} &
    \includegraphics[height=5cm]{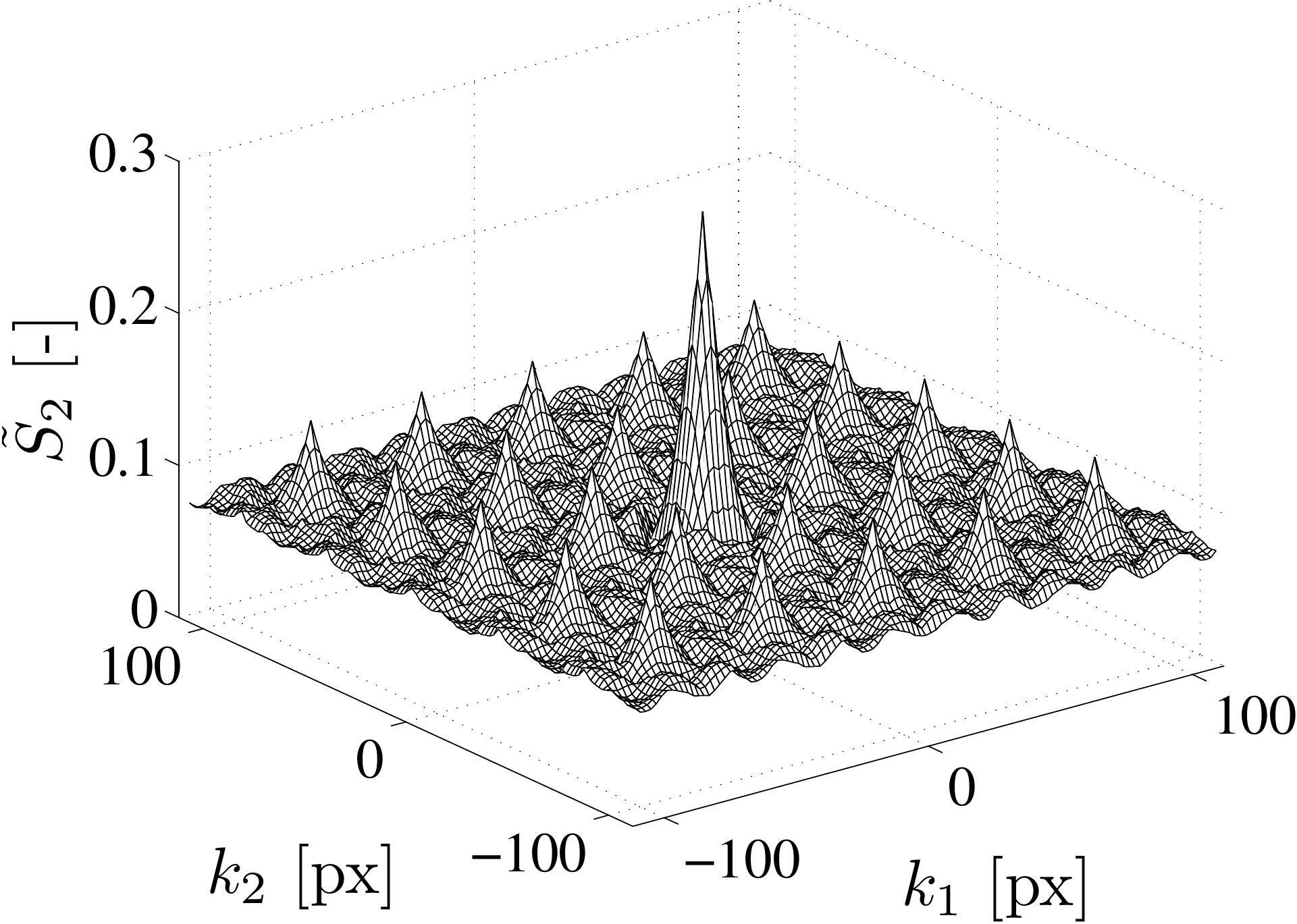} \\
    (b) &
    \includegraphics[height=5cm]{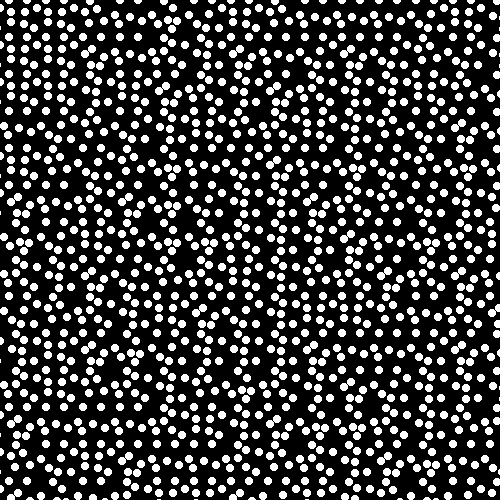} &
    \includegraphics[height=5cm]{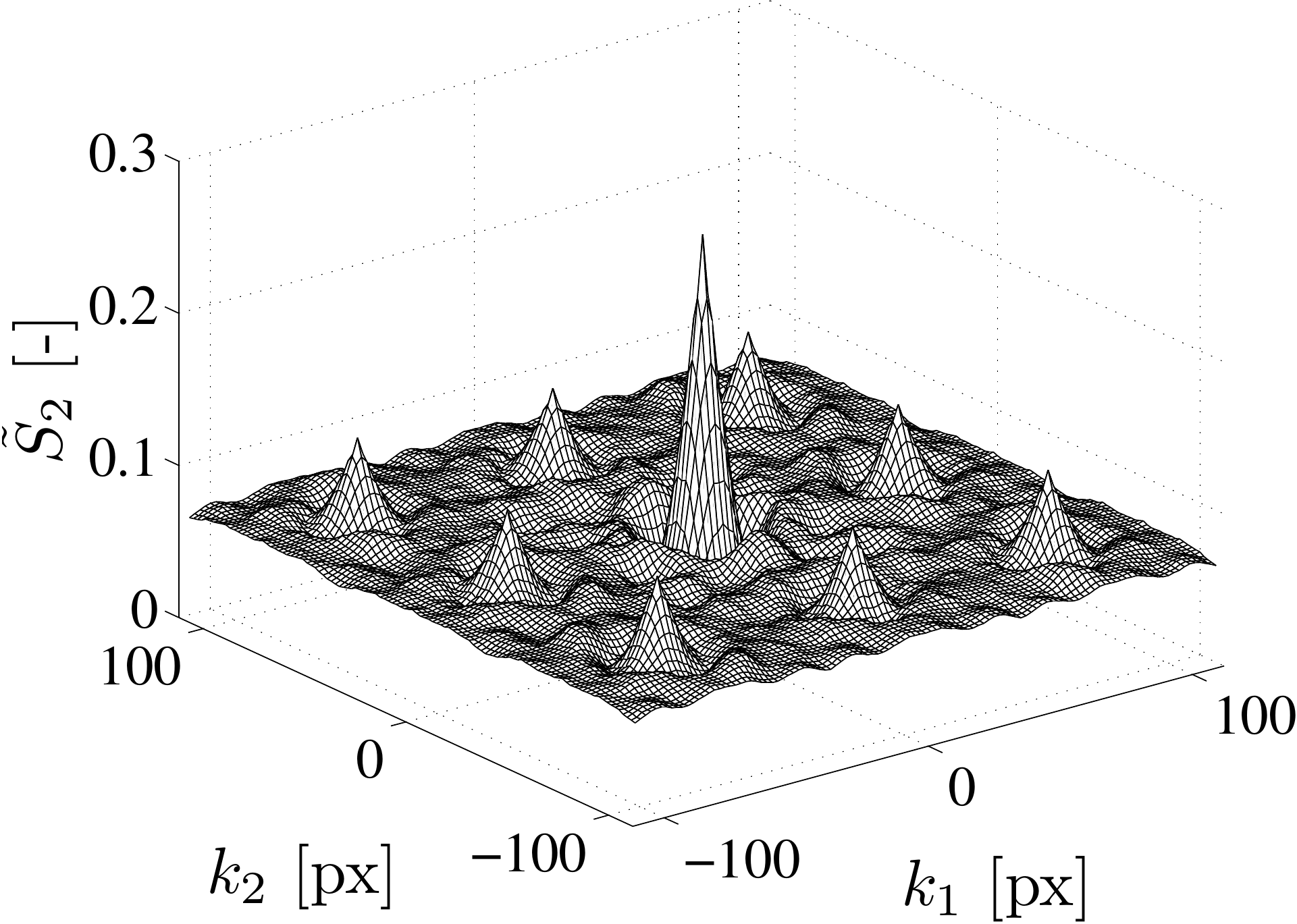}\\
    (c) &
    \includegraphics[height=5cm]{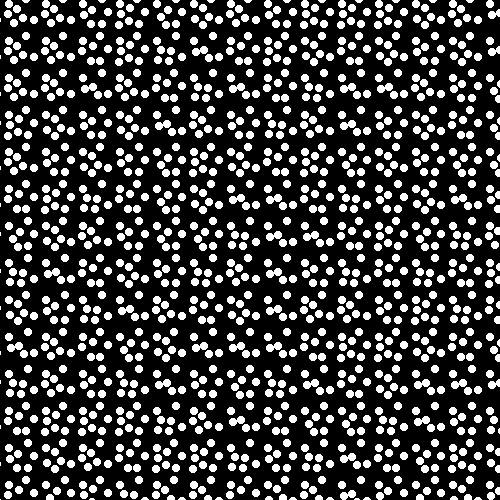} &
    \includegraphics[height=5cm]{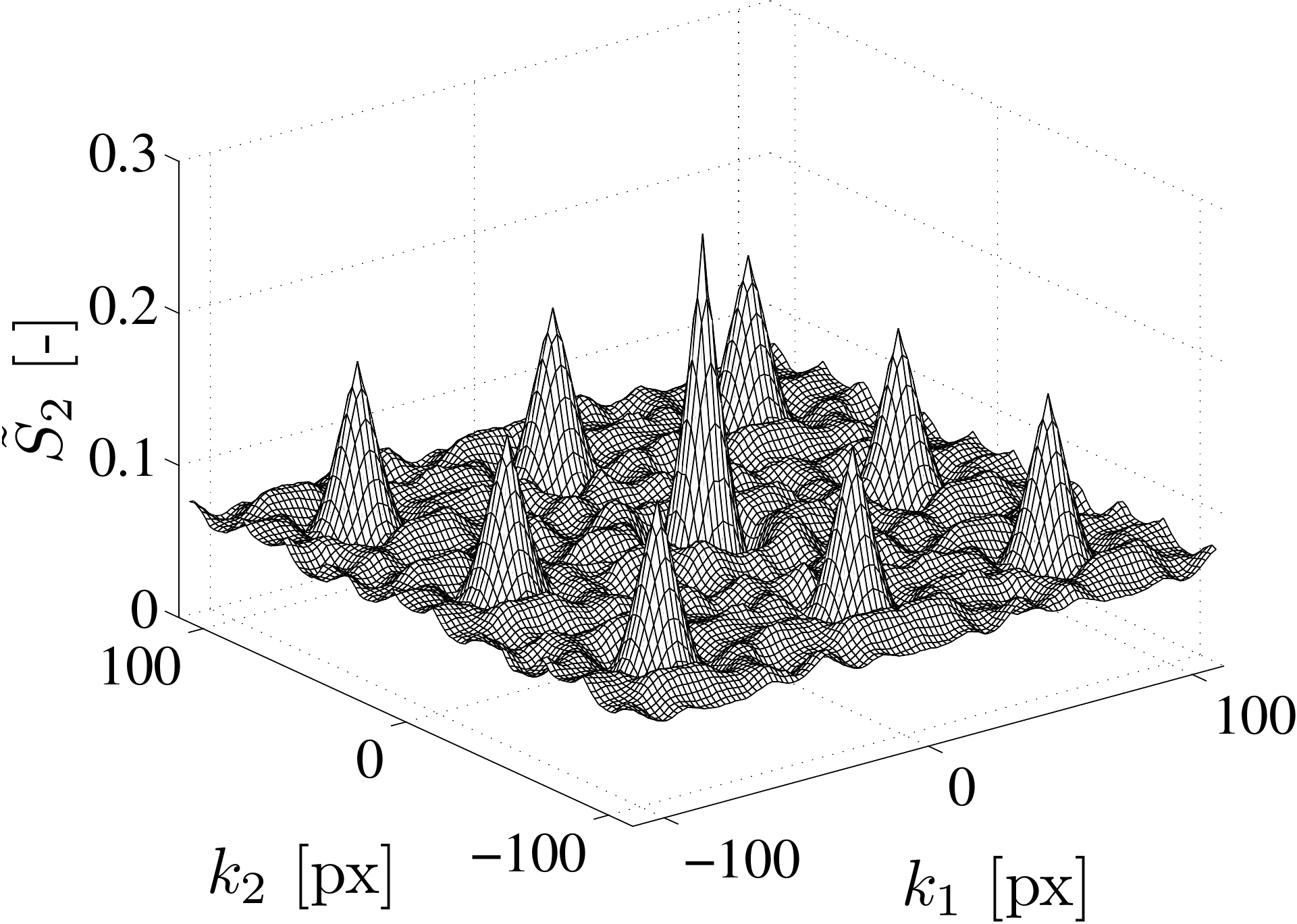}
    \end{tabular}
  \caption{Reconstructed microstructures and two point probability functions
  $\rS$ for tile sets with (a)~$\wf = 10^5, \np=10$ disks and $\ml=42$~px,
  (b)~$\wf = 10^5, \np=38$ disks and $\ml=74$~px, and 
  (c)~$\wf = 10^4, \np=38$ disks and $\ml=74$~px.}
  \label{fig:results_S2} 
\end{figure}

In \Fref{fig:results_S2}, we present the disk configurations and two-point
probability functions $\rS$ obtained for the domain $\dmn$ being tiled by
optimised tile sets. We observe that all reconstructed functions $\rS$ exhibit
local peaks exceeding the value of $\vfrac^2$, which reveals the presence of
characteristic length scales of order $\ml$ in the synthesised medium.
\complete{For the default value of the weighting factor $\wf = 10^5$,
Figs.~\ref{fig:results_S2}(a,b)}, the local extremes are notably smaller than
the value of $\vfrac$ corresponding to a periodic construction,
e.g.~\cite{Zeman:2007:FRM}. In addition, their number \rev{and magnitude} can be
substantially reduced by increasing the edge length $\ml$,
\Fref{fig:results_S2}(b), \rev{and practically eliminated when using more
general tile sets~\cite{Novak:2012:CRM}.} \complete{For lower values of $\wf$,
the disk distribution becomes more regular, \Fref{fig:results_S2}(c), and the
resulting representation is visually indistinguishable from the periodic
setting, cf.~\cite{Novak:2012:CRM}.}

\begin{figure}[ht]
  \centering
  \begin{tabular}{cc}
    (a)\includegraphics[width=.45\textwidth]{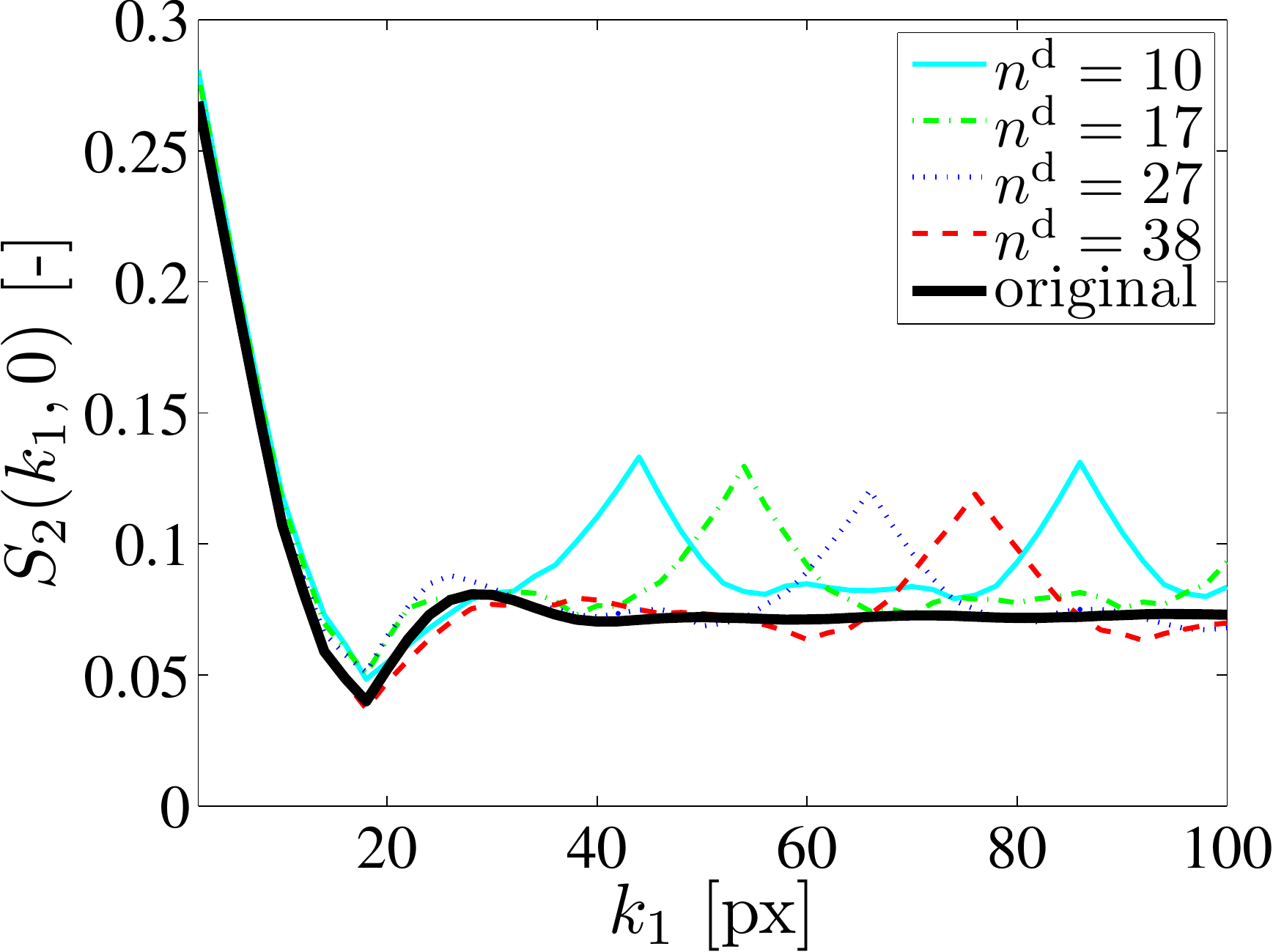}
    &
    (b)\includegraphics[width=.45\textwidth]{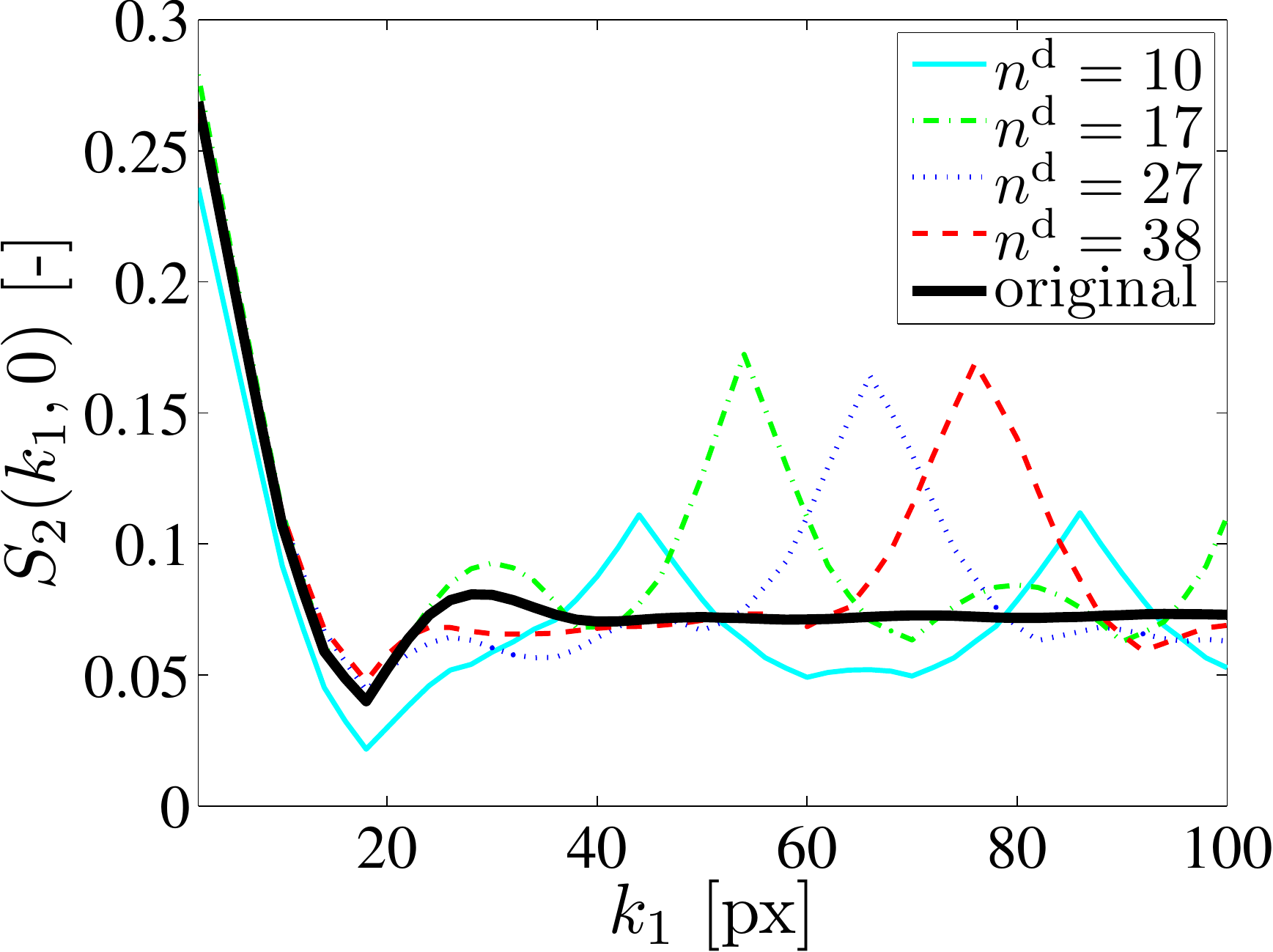}
    \end{tabular}
  \caption{Comparison of two-point probability functions $\SII(k_1, 0)$ for the
  weighting factors (a)~$\wf=10^5$ and (b)~$\wf=10^4$.}
\label{fig:alpha_acf_cuts} 
\end{figure}

Such conclusions are further supported by \Fref{fig:alpha_acf_cuts} showing
cross-sections of the two-point probability functions in the $k_1$ direction for
two different values of the weighting factor $\wf$. The results demonstrate
that for higher values of $\wf$, the short-range phenomena are captured to a high
accuracy and the magnitude of local extremes are consistently reduced with the
increasing number of disks, albeit at a small rate. By decreasing the emphasis
on $\SII$ objective, \Fref{fig:alpha_acf_cuts}(b), the discrepancy
between the original and reconstructed medium substantially increases at short
distances, leading even to an inconsistent value of the volume fraction for $10$
disks. The local peaks also become more pronounced as the stress-based criterion
drives the system towards periodic configurations.

\rev{%
\Fref{fig:traction_optimization}(a) illustrates the ability of the optimisation
algorithm to achieve self-equlibrated stress enrichment functions by comparing
the distribution of tractions $\Tfl{1}{3}$ obtained for an initial and the
optimised configuration of disks. Clearly, traction enrichments at contiguous
edges differ significantly in the initial configuration, and are reduced to
almost identical values by the proposed procedure. \complete{This also
automatically keeps the edge jumps in reconstructed traction enrichments
$\jmp{\rTfl{1}{3}}$ under control, \Fref{fig:traction_optimization}(b), since
their magnitude corresponds to the scatter found for representative eight tiles
from $\dmnT$ utilised in the reconstruction, recall
\Fref{fig:optimproc}(b).}
\begin{figure}[ht]
  \centering
  \begin{tabular}{cc}
    (a)\includegraphics[width=.45\textwidth]{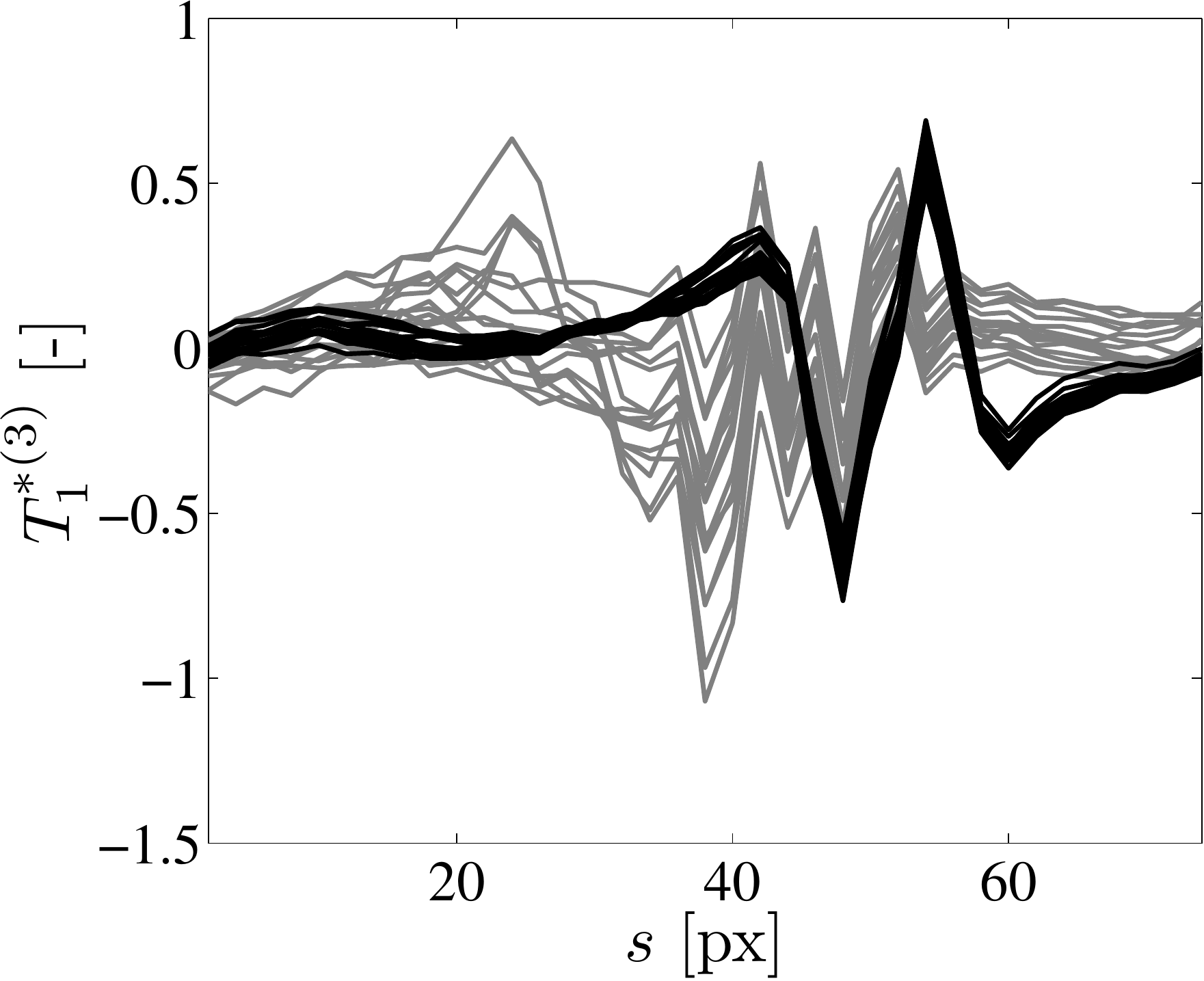}
    &
    (b)\includegraphics[width=.45\textwidth]{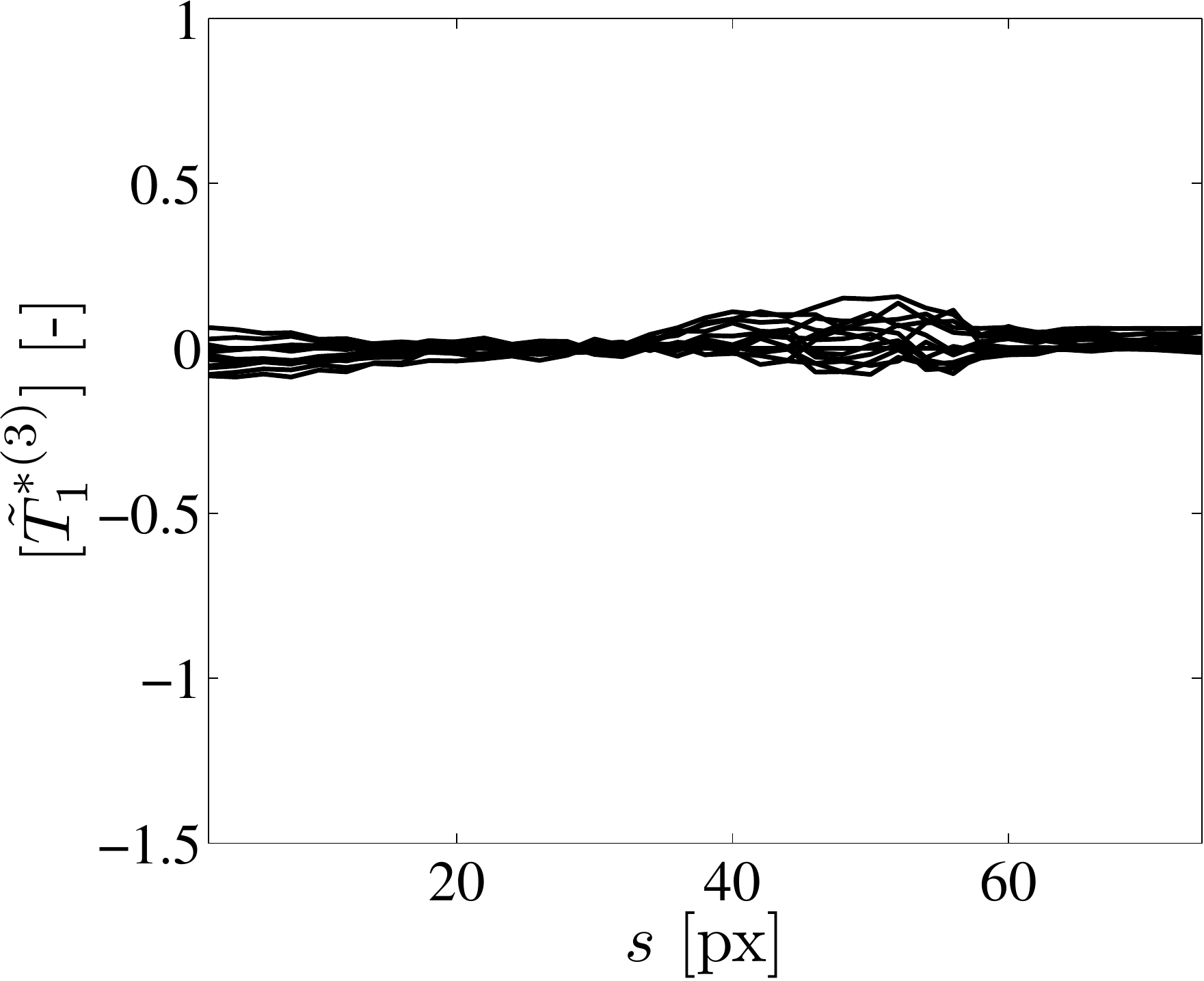}
    \end{tabular}
  \caption{Distribution of (a)~traction enrichments $\Tfl{1}{3}$ and
  (b)~reconstructed traction enrichment jumps $\jmp{\rTfl{1}{3}}$ at edges
  $\Ge{\delta}$ of the tiling $\dmnT$, obtained for $\np = 38$ disks and $\ml =
  74$~px. The grey/black patterns in (a) correspond to initial and optimised and
  enrichment functions, respectively.}
  \label{fig:traction_optimization}
\end{figure}

\complete{To what extent influences such choice of representative tiles 
the synthesised enrichment functions?} To address this question, we consider a
particular reconstruction of stress enrichments functions $\rec{\MS}\flc : \dmnT \rightarrow \R^{3 \times 3}$,
assembled according to the the sequence of tiles found in the tiling $\dmnT$. It
is useful for the visualisation purposes to introduce a local error measure
\begin{align}
\fSigloc{ij}(\k)
=
\frac{
\left| \Sigma\flc_{ij}(\k) - \rec{\Sigma}\flc_{ij}(\k) \right|
}{%
\displaystyle
\max_{\m \in \ZdmnT} \Sigma\flc_{ij}(\m)
-
\min_{\m \in \ZdmnT} \Sigma\flc_{ij}(\m)
},
&&
\k \in \ZdmnT;
i,j \in \{ 1, 2, 3 \},
\end{align}
\nomenclature{$\fSigloc{ij}$}{Local stress-based objective function}%
\nomenclature{$\fSig$}{Global stress-based objective function}%
quantifying a difference between the components of the stress enrichment
functions ${\MS}\flc$ determined directly for the tiling $\dmnT$ by the
algorithm described in \Aref{app:mech_fields}, and their reconstruction
$\rec{\MS}\flc$. 

\begin{figure}[ht]
{\small
  \begin{tabular}{cccc}
    \multicolumn{4}{c}{$\np=10$ disks, $\wf = 10^5$} \\
    \includegraphics[width=3.35cm]{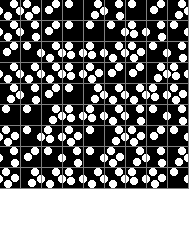}
    &\includegraphics[width=3.35cm]{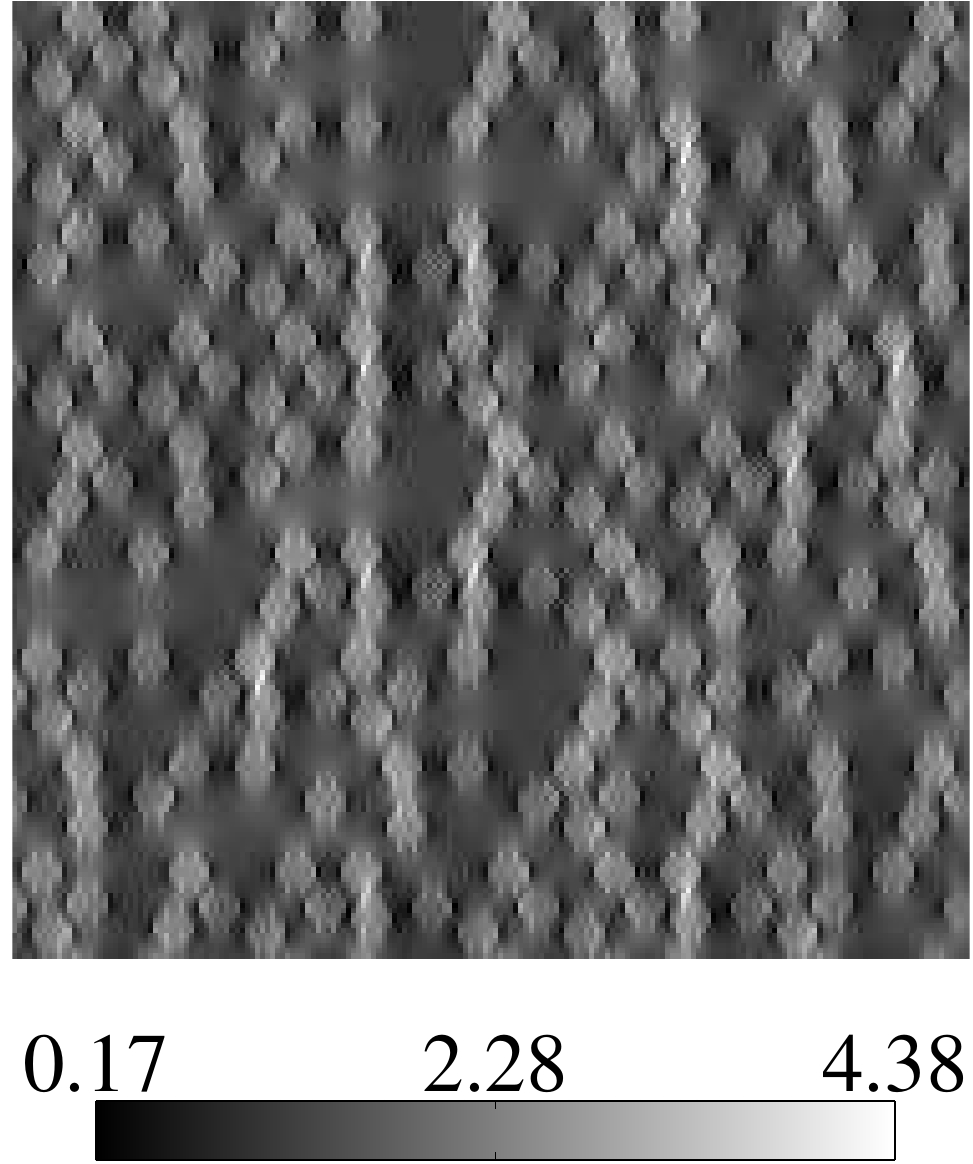}
    &\includegraphics[width=3.35cm]{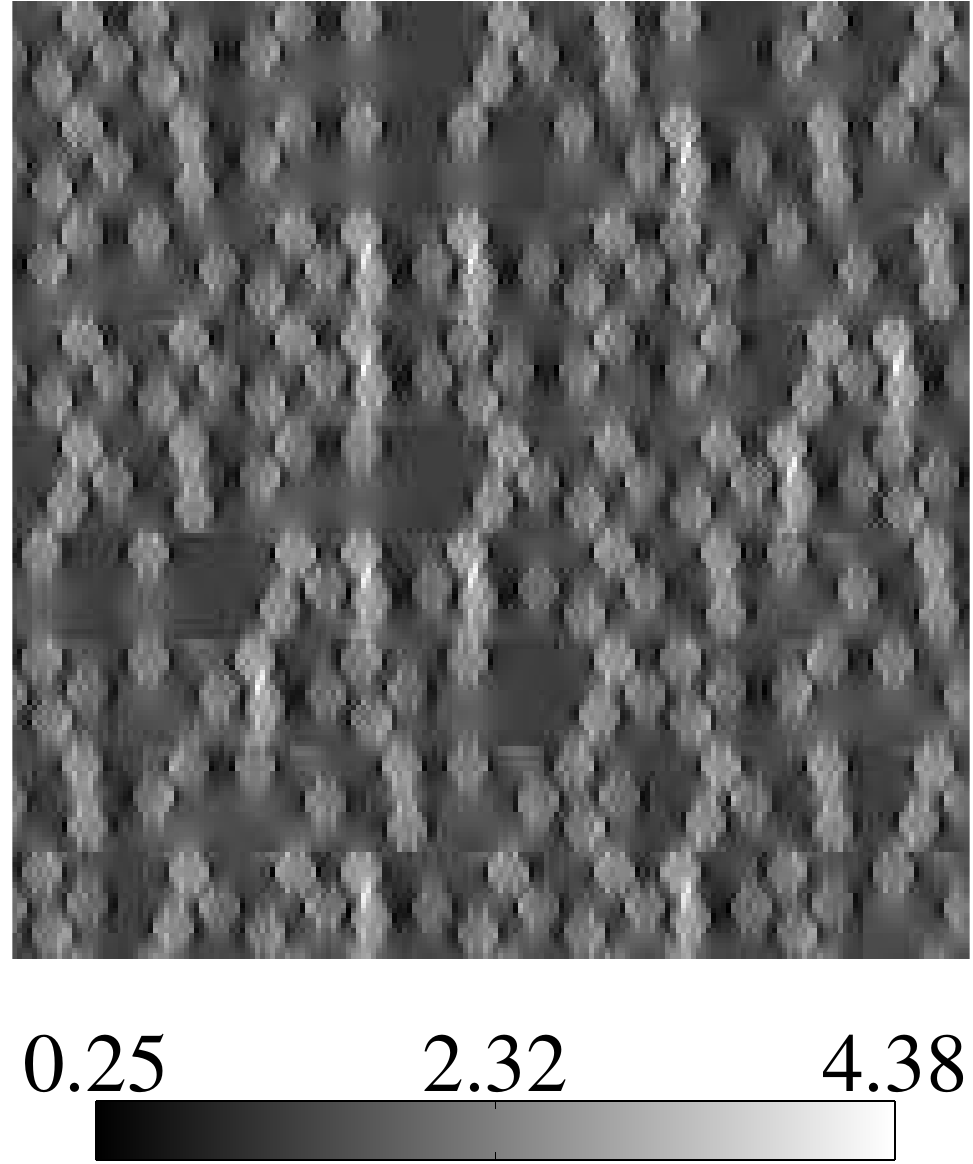}
    &\includegraphics[width=3.35cm]{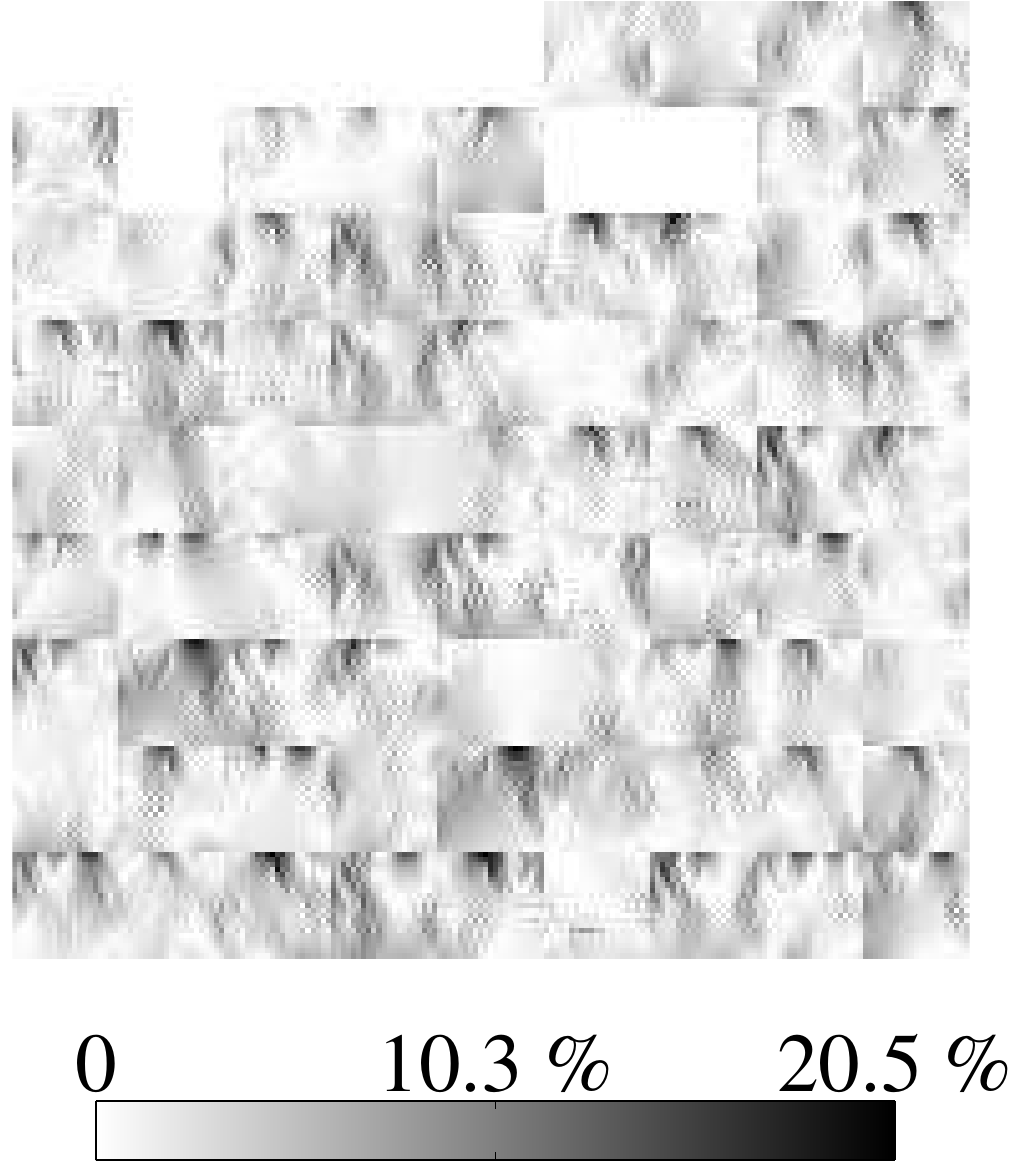}\\
    \multicolumn{4}{c}{$\np = 38$ disks, $\wf = 10^5$} \\
    \includegraphics[width=3.35cm]{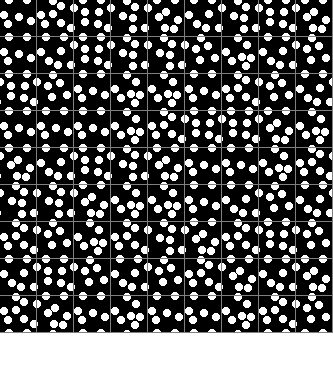}
    &\includegraphics[width=3.35cm]{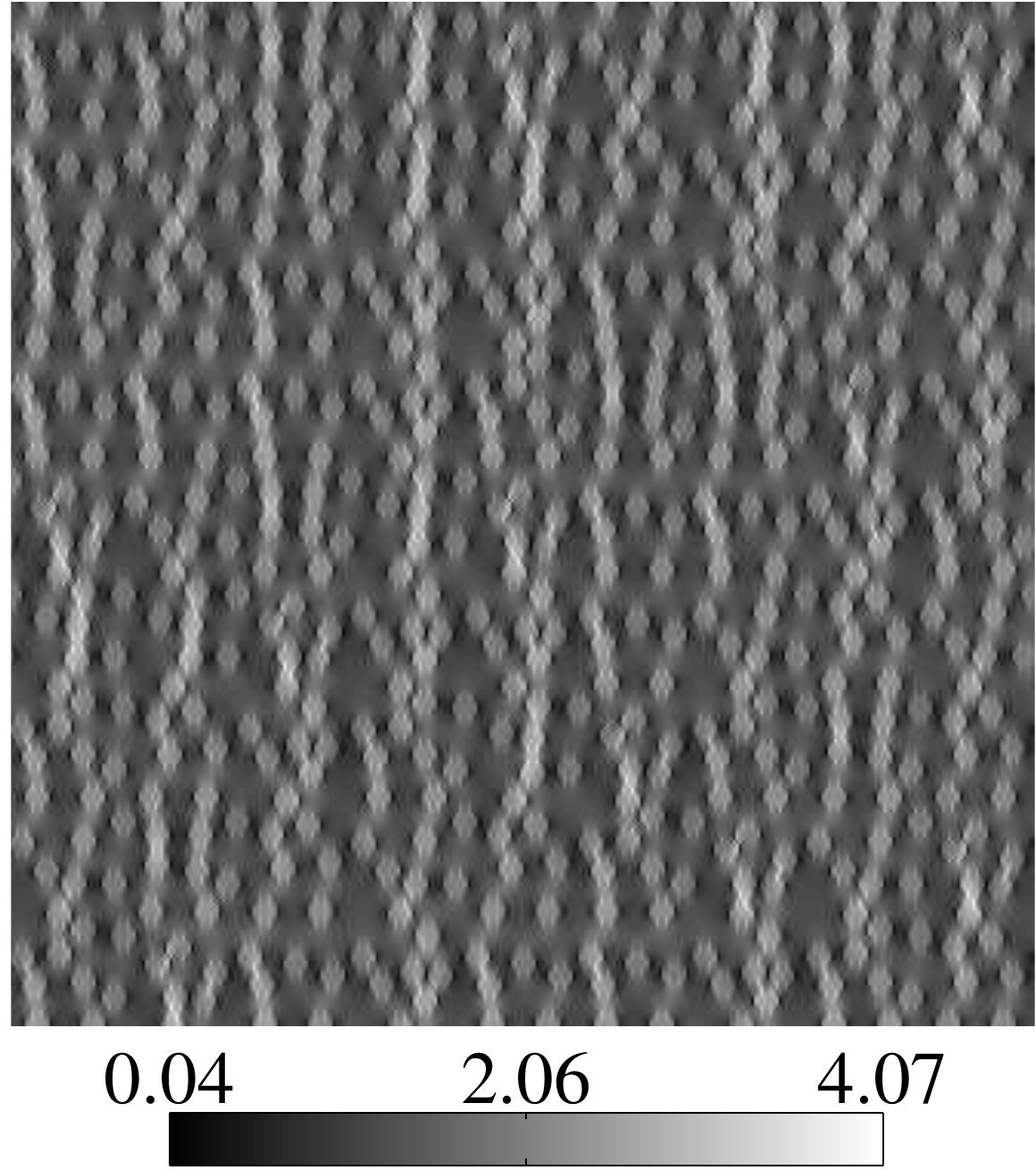}
    &\includegraphics[width=3.35cm]{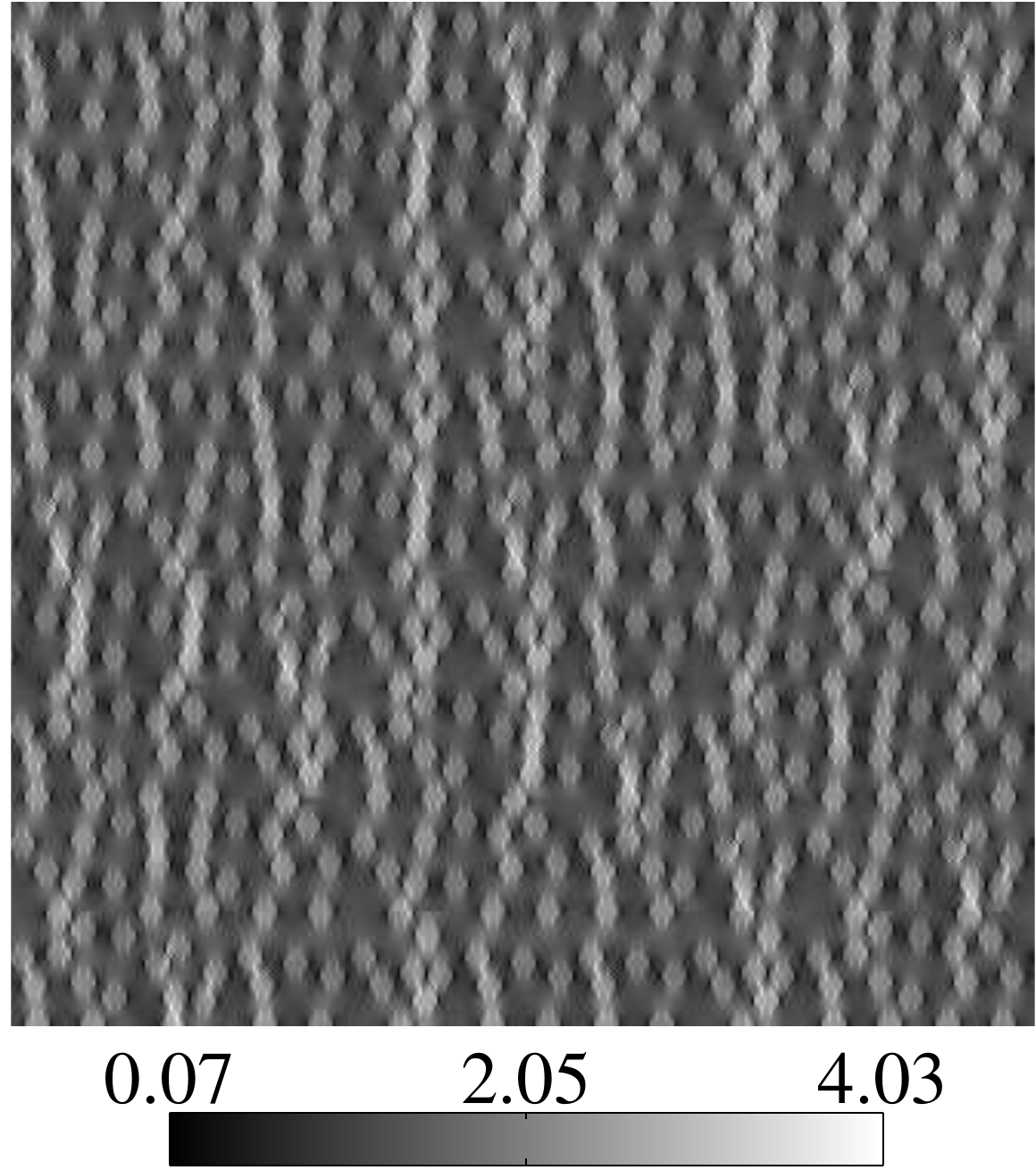}
    &\includegraphics[width=3.35cm]{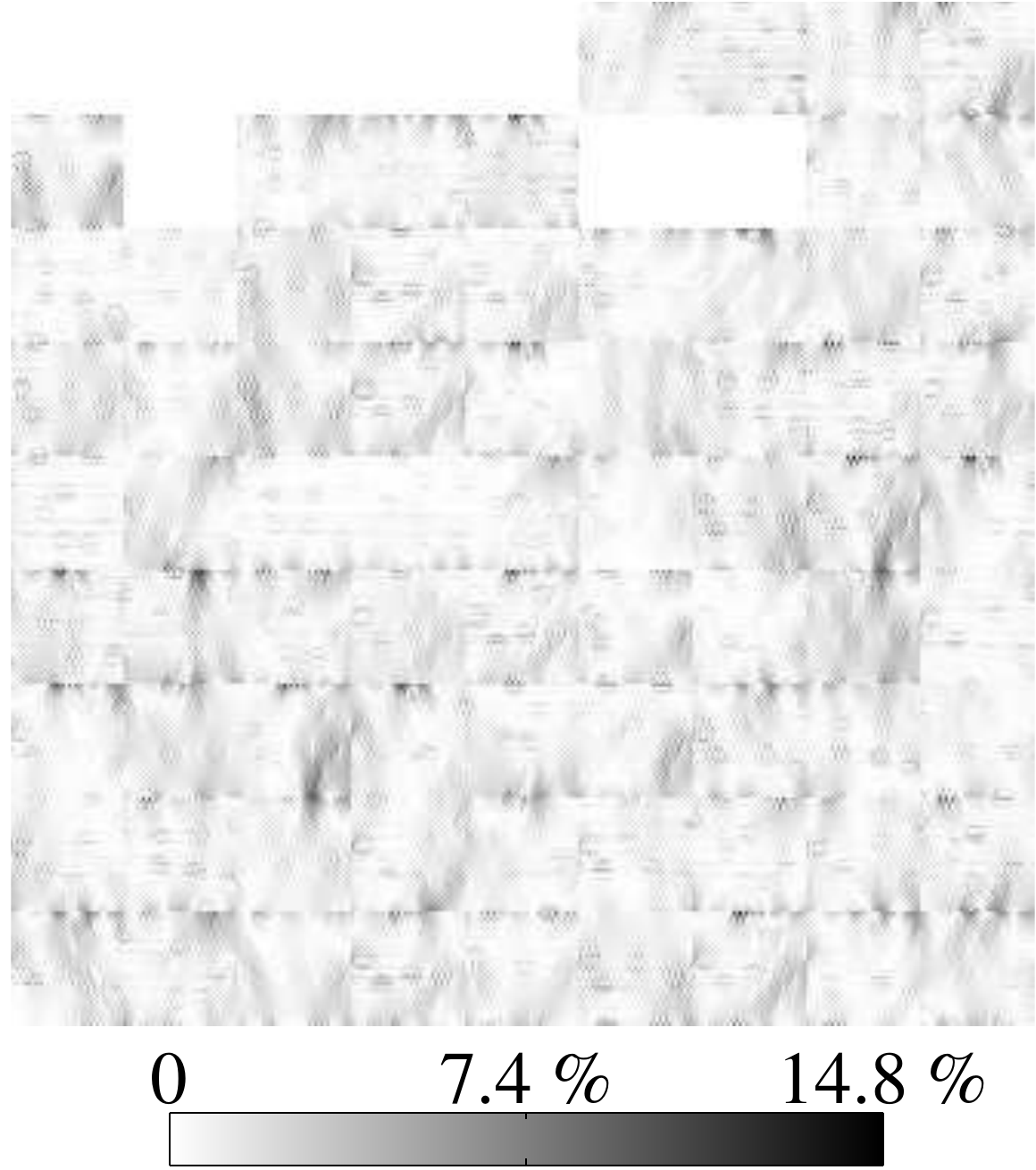}\\
    (a) &(b) &(c) &(d)
    \end{tabular}
}
  \caption{Assessment of tiling-based enrichment functions,
  (a)~microstructures obtained by tilings $\dmnT$, distribution of (b)~true
  stress enrichment functions $\Sigma\flc_{13} \equiv \sigfl{11}{3}$,
  (c)~reconstructed stress enrichment functions $\rec{\Sigma}\flc_{13} \equiv
  \rsigfl{11}{3}$ and of (d)~the local reconstruction-based error
  $\fSigloc{13}$.
\label{fig:reconstructed_stress_fields} 
}
\end{figure}

Outcomes of this comparison are shown in~\Fref{fig:reconstructed_stress_fields}
in the form of (a)~tiling-based microstructures, (b)~distribution of the
corresponding enrichment functions $\Sigma\flc_{13}$, (c)~their reconstructed
counterparts and (d)~spatial distribution of the relative error. For the
microstructure generated from tiles with $\np=10$ disks, we observe that the
reconstructed field displays distributed errors in tile interiors.
Similarly to $\SII$ criterion, these deviations are significantly reduced
and become highly localised when increasing the number of disks and the size of
tiles. This claim is further supported by~\Fref{fig:global_error_evolution},
plotting the evolution of the global error
\begin{equation}
\fSig 
= 
\frac{1}{\dmnT} 
\sum_{i,j=1}^3
\sum_{\k \in \ZdmnT}
| \fSigloc{ij}(\k) |
\end{equation}
as a function of the number of disks. For both values of $\wf$, we observe
approximately linear convergence with increasing $\np$. In addition, the error
decreases for larger phase contrasts $\Ed/\Em$. This is caused by the
fact that stresses tend to concentrate more at stiffer disks, therefore reducing
variations of tractions at tile edges, see also~\cite{Novak:2012:MEF} for a
similar discussion.

\begin{figure}[ht]
  \centering
  \begin{tabular}{cc}
    (a)\includegraphics[width=.45\textwidth]{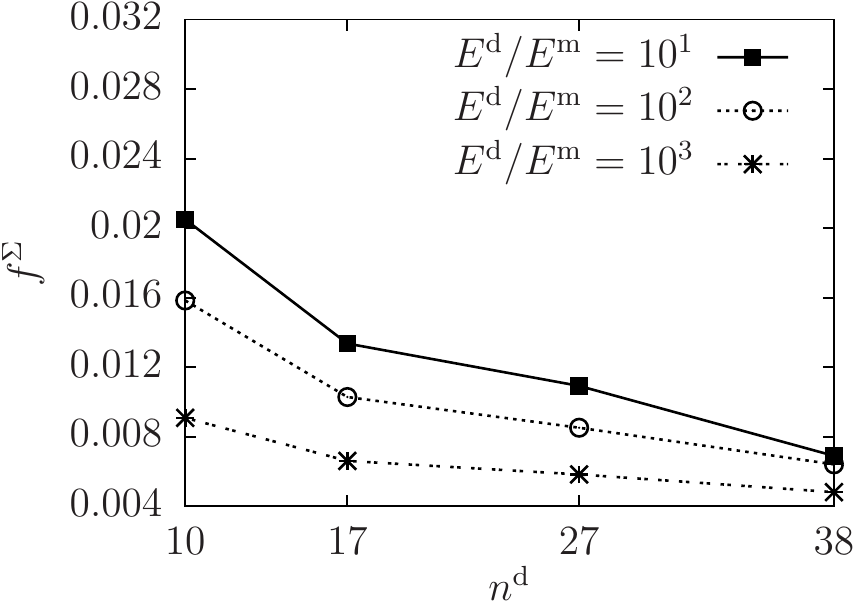}
    &
    (b)\includegraphics[width=.45\textwidth]{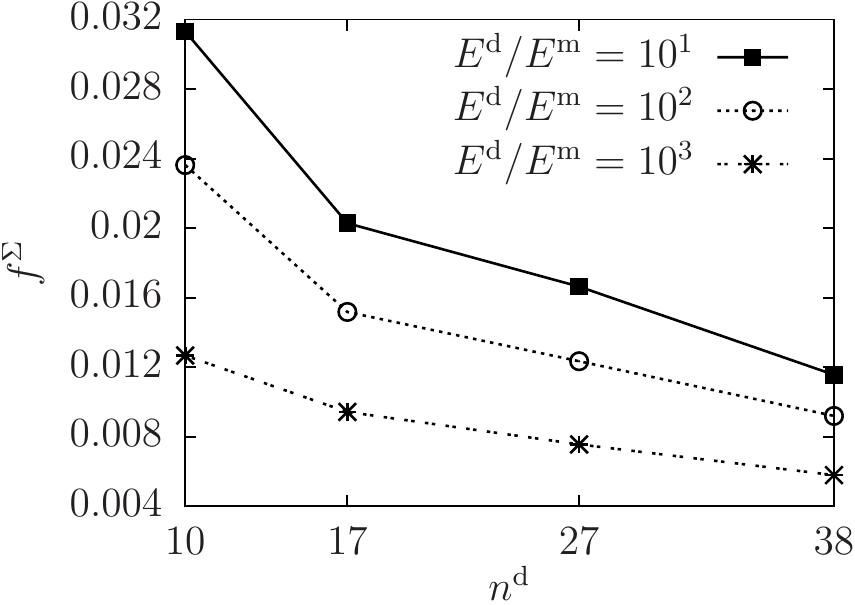}\\
    \end{tabular}
  \caption{The global reconstruction error $\fSig$  as a function of the number
  of disks $\np$ for different phase contrasts $\Ed/\Em$ and weighting factors
  (a)~$\wf=10^5$ and (b)~$\wf=10^4$.}
\label{fig:global_error_evolution}
\end{figure}

Altogether, this indicates that the tile set was designed correctly, since the
optimisation was executed for independent objective functions, recall
\Eref{eq:err_tot}. Finally we stress that the significant compression has been
achieved by the tiling-based representation: the original microstructure
contains $\approx 1,300$ disks, whereas the most detailed tile-based
representation builds on $38$ disks only and is capable of producing much larger
microstructures at a negligible computational cost. 
}

\section{Conclusions}\label{s:conclusions}
In this work, we have proposed an approach to the construction of aperiodic
local fields in heterogeneous media with potential applications in hybrid or
generalised FE environments. The method is based on the Wang tiling concept that
allows us to represent complex patterns using a limited set of representative
tiles, complemented by the Simulated Annealing-based algorithm to arrive at 
optimal tile set morphologies. On the basis of the results obtained from
analyses of the medium under consideration we conjecture that:

\begin{itemize}
\item the proposed method provides a robust tool for compression of disordered
microstructures and can serve as an efficient microstructure generation
algorithm, 
\item it allows for aperiodic extensions of local, possibly periodic,
fields to substantially larger domains while maintaining their compatibility,
\item the tiling-based fields can be utilised as
microstructure-based enrichment functions for generalised Partition of Unity methods or hybrid finite
element schemes.
\end{itemize}

We are fully aware that our conclusions are somewhat provisional, in the sense
that these are based on a single set of tiles and the specific class of
microstructures. Partial extension to general setting is available
in~\cite{Novak:2012:CRM,Doskar:2012:GMH} and remains in the focus of our current work.

\paragraph{Acknowledgements}
The authors thank Jaroslav Vond\v{r}ejc~(CTU in Prague) for providing us with a
MATLAB source code of FFT-based homogenisation algorithm and Adrian
Russell~(University of New South Wales), Michal \v{S}ejnoha and Milan
Jir\'{a}sek~(CTU in Prague) and anonymous referees for their criticism and
helpful comments on earlier versions of the manuscript. We also gratefully
acknowledge financial support by the Czech Science Foundation through grants
No.~P105/12/0331~(JN), P105/11/P370~(AK), and P105/11/0411~(JZ). Our work was
partially supported by the European Social Fund, grant
No.~CZ.1.07/2.3.00/30.0005 of Brno University of Technology (Support for the
creation of excellent interdisciplinary research teams at Brno University of
Technology, JN), by the Ministry of Education, Youth and Sports of the Czech
Republic through project MSM~6840770003~(AK), and by the European Regional
Development Fund under the IT4Innovations Centre of Excellence, project
No.~CZ.1.05/1.1.00/02.0070~(JZ).

\appendix

\section{Computation of mechanical fields}\label{app:mech_fields}

As explained earlier in \Sref{sec:introduction}, our objective is to
determine local fields within a given domain $\puc \subset \R^2$
subjected to a given overall strain field 
\begin{equation}\label{eq:overall_strain}
\MEps
=
\begin{bmatrix}
\Eps_{11} & \Eps _{22} & \sqrt{2} \Eps_{12}
\end{bmatrix}\trn,
\end{equation}
under the periodic boundary conditions. These follow from the solution of the
elastic unit cell problem~\cite{Milton:2002:TC,Michel1999109}
\nomenclature{$\puc$}{Arbitrary periodic domain}%
\nomenclature{$\Eps$}{Macroscopic strain field}%
\begin{eqnarray}\label{eq:unit_cell_problem}
\Meps(\x)
= 
\Mdiff
\Mu(\x),
&
\Mdiff\trn
\Msig(\x)
=
\M{0},
&
\Msig(\x)
=
\ML(\x)
\Meps(\x)
\mbox{ for }
\x \in \puc,
\end{eqnarray}
in which $\Mu : \puc \rightarrow \R^2$ designates the displacement field,
$\Meps : \puc \rightarrow \R^3$ and $\Msig : \puc \rightarrow \R^3$ denote the
$\puc$-periodic strain and stress fields, $\ML : \puc \rightarrow \R^{3 \times
3}$ stands for the symmetric positive-definite material stiffness matrix, and
the operator matrix is defined as
\begin{equation}
\Mdiff
=
\begin{bmatrix}
\frac{\partial}{\partial x_1} & 0 & 
\frac{1}{\sqrt 2}\frac{\partial}{\partial x_2}
\\
0 & \frac{\partial}{\partial x_2} & 
\frac{1}{\sqrt 2}\frac{\partial}{\partial x_1}
\end{bmatrix}\trn.
\end{equation}
\nomenclature{$\eps$}{Strain}%
\nomenclature{$\ML$}{Matrix of material stiffness}%
\nomenclature{$\Mdiff$}{Operator matrix}%
In addition, the strain field is subject to a mean value-type constraint
\begin{equation}
\frac{1}{|\puc|}
\int_{\puc}
\Meps( \x )
\de \x
=
\MEps.
\end{equation}

It is well-known~\cite{Michel1999109,Milton:2002:TC} that the solution to the
unit cell is characterised by the Lippmann-Schwinger equation
\begin{equation}
\Meps( \x )
+
\int_{\puc}
\Mgfune(\x - \y )
\delta \ML( \y )
\Meps( \y )
\de \y
=
\MEps
\mbox{ for }
\x \in \puc,
\end{equation}
\nomenclature{$\Mgfune$}{Green function of the reference medium}%
where $\delta \ML = \ML - \ML^0$, $\ML^0 \in \R^{3\times 3}$ is the stiffness
matrix of an auxiliary reference medium and the operator $\Mgfune : \puc
\rightarrow \R^{3\times 3}$ is related to the Green function of the
problem~\eqref{eq:unit_cell_problem} with $\ML(\x) = \ML^0$. It admits a compact
closed-form expression in the Fourier space, e.g.~\cite[Section
5.3]{Michel1999109}, and its action can be efficiently evaluated by the FFT
algorithm. This observation is at the heart of an iterative scheme due to
Moulinec and Suquet~\cite{Moulinec:1994:FNMC}, which can be applied to arbitrary
digitised media.

In our case, we adopt an accelerated version of the original algorithm based on
observations due to Zeman \emph{et al.}~\cite{Zeman:2010:AFFT}. Since the sample
is discretized by a regular $\npuc_1 \times \npuc_2$ bitmap, it is
convenient to project the integral equation onto the space of trigonometric
polynomials, e.g.~\cite{Saranen:2002:PIP}. This yields the linear system in the
form
\nomenclature{$\npuc_i$}{Number of pixels for $\puc$ in the $i$-th direction}%
\begin{equation}\label{eq:FFT_system}
( \M{I} + \M{B} ) \M{e} = \M{b},
\end{equation}
where $\M{e} \in \set{R}^{3\times \npuc_1 \times \npuc_2}$ stores
the unknown strain values at individual pixels, $\M{b} \in \set{R}^{3 \times
\npuc_1 \times \npuc_2}$ is the corresponding matrix of overall
strains and matrix $\M{B}$ is expressed as a product of several matrices
\begin{equation}
\begin{split}
\M{B}
& = 
\begin{pmatrix}
\M{F}^{-1} & \M{0} & \M{0} \\
\M{0} & \M{F}^{-1} & \M{0} \\
\M{0} & \M{0} & \M{F}^{-1} 
\end{pmatrix}
\begin{pmatrix}
\M{\Gamma}_{1111} & 
\M{\Gamma}_{1122} & 
\sqrt{2} \M{\Gamma}_{1112} \\
\M{\Gamma}_{2211} & 
\M{\Gamma}_{2222} & 
\sqrt{2} \M{\Gamma}_{1112} \\
\sqrt{2}\M{\Gamma}_{2212} & 
\sqrt{2}\M{\Gamma}_{1212} & 
2 \M{\Gamma}_{2212} 
\end{pmatrix}
\begin{pmatrix}
\M{F} & \M{0} & \M{0} \\
\M{0} & \M{F} & \M{0} \\
\M{0} & \M{0} & \M{F} 
\end{pmatrix}
\\
& \times 
\begin{pmatrix}
\delta\M{L}_{1111} & 
\delta\M{ L}_{1122} & 
\sqrt{2} \delta\M{ L}_{1112} \\
\delta\M{ L}_{2211} & 
\delta\M{ L}_{2222} & 
\sqrt{2} \delta\M{ L}_{1112} \\
\sqrt{2}\delta\M{ L}_{2212} & 
\sqrt{2}\delta\M{ L}_{1222} & 
2 \delta\M{ L}_{2212} 
\end{pmatrix}.
\end{split}
\end{equation}
Here, $\M{F} \in \set{C}^{\npuc_1 \times \npuc_2}$ and $\M{F}^{-1}$
implement the forward and the inverse Fourier transform and, e.g., $\delta\M{
L}_{1122} \in \set{R}^{\npuc_1 \times \npuc_2}$ stores the
corresponding component of the stiffness tensor at individual pixels,
see~\cite{Zeman:2010:AFFT} for more details. The system~\eqref{eq:FFT_system} is
solved using standard conjugate gradient algorithm.
Upon convergence, the distribution of the local stress field $\Msig$
is determined from the solution $\M{e}$ by \Eref{eq:unit_cell_problem}$_3$. The
local displacement fields $\Mu$ follow from an inexpensive analysis in the
Fourier space, e.g.~\cite{Willot:2008:EMT}. 

Note that the construction of the enrichment functions is
based on the perturbation fields of displacements and stresses
\begin{eqnarray}\label{eq:fluctuating_fields}
\Mu\flc(\x)
& =
\Mu( \x ) 
-
\frac{1}{|\puc|}
\int_{\puc}
\Mu( \y ) 
\de\y,
\\
\Msig\flc(\x)
& =
\Msig( \x ) 
-
\frac{1}{|\puc|}
\int_{\puc}
\Msig( \y )
\de\y,
\end{eqnarray}
instead of the total values. \rev{ The enrichment functions for displacements,
$\MU\flc$ in \Eref{eq:displ_enrichments}, and stresses, $\MS\flc$
in~\Eref{eq:stress_enrichments}, can now be constructed from the solutions to
three load-cases, obtained by successively setting each component of
$\Eps_{ij}$ in~\eqref{eq:overall_strain} to $1$, while the ones become $0$.}

\end{document}